\documentclass[prd,twocolumn,english,preprintnumbers,amsmath,amssymb,nofootinbib,a4]{revtex4}

\usepackage[latin1]{inputenc}
\usepackage{graphicx}
\usepackage{color}
\usepackage{bbm}
\usepackage{amssymb}
\usepackage{amsmath}

\usepackage{dsfont}

\def\0#1#2{\frac{#1}{#2}}

\def\s0#1#2{\mbox{\small{$ \frac{#1}{#2} $}}}

\newcommand{\tr}{\mathrm{tr}}
\newcommand{\trf}{\mathrm{tr}_{\rm F}}
\newcommand{\E}{\mathrm{e}}
\newcommand{\I}{\mathrm{i}}
\newcommand{\be}{\begin{eqnarray}}
\newcommand{\ee}{\end{eqnarray}}

\newcommand{\nn}{\nonumber }
\newcommand{\fslash}{\hspace*{-0.2cm}\slash }

\newcommand{\Nf}{N_{\text{f}}}
\newcommand{\Nc}{N_{\text{c}}}

\newcommand{\bfe}{\langle A_0\rangle}
\newcommand{\Td}{T_{\rm d}}
\newcommand{\Tc}{T_{\chi}}
\newcommand{\luv}{\lambda_{\psi}^{\rm UV}}
\newcommand{\lfp}{\lambda_{\psi}^{\ast}}

\usepackage{babel}
\makeatother

\begin{document}

\title{Dynamical Locking of the Chiral and the Deconfinement Phase Transition in QCD}
\author{Jens Braun} 
\author{Alexander Janot}
\affiliation{Theoretisch-Physikalisches Institut,
  Friedrich-Schiller-Universit\"at Jena, D-07743 Jena, Germany}

\begin{abstract}
We study the fixed-point structure of four-fermion interactions in two-flavor QCD with 
$\Nc$ colors close to the finite-temperature phase boundary. In particular, we 
analyze how the fixed-point structure of four-fermion interactions is related 
to the confining dynamics in the gauge sector. We show that there  
exists indeed a mechanism which dynamically locks the chiral phase transition 
to the deconfinement phase transition. This mechanism allows us to 
determine a window for the values of physical observables 
in which the two phase transitions lie close to each other.
\end{abstract}

\maketitle

\section{Introduction}
The relation of quark confinement and chiral symmetry breaking in quantum 
chromodynamics (QCD) is not yet fully understood. In particular at finite temperature
and quark chemical potential the investigation of the QCD phase boundary represents one of the 
major research topics in theoretical physics and is also of great importance for a
better understanding of heavy-ion collision experiments~\cite{BraunMunzinger:2003zd}. 
Theoretical studies in the limit of many colors in fact suggest that an understanding 
of the interrelation of the chiral and the deconfinement phase transition is required to 
comprehend the QCD phase structure close to a possible critical endpoint~\cite{McLerran:2007qj}.

While the confinement transition is driven by the gauge degrees of freedom,
the chiral phase transition is triggered by strong quark self-interactions.
In Nambu--Jona-Lasinio (NJL) models (and in the Polyakov-loop extended version
thereof) these quark interactions are considered as parameters~\cite{Meisinger:1995ih,Pisarski:2000eq,%
Mocsy:2003qw,Fukushima:2003fw,Megias:2004hj,Ratti:2005jh,Sasaki:2006ww,Schaefer:2007pw,Mizher:2010zb,Skokov:2010wb,Herbst:2010rf,Skokov:2010uh}
and used to fit a given set of low-energy observables.
In full QCD, however, we expect that the quark self-interactions are dynamically generated and driven to criticality 
by the gauge degrees of freedom. This has indeed been confirmed by means of a renormalization group (RG)
analysis of the influence of gluodynamics on the fixed-point structure of four-fermion interactions, see e.~g. 
Refs.~\cite{Gies:2002hq,Gies:2005as,Braun:2005uj,Braun:2006jd,Braun:2008pi,Braun:2009ns,Braun:2010qs}: 
once the gauge coupling exceeds a critical value, the quark sector is driven to criticality without 
requiring any fine-tuning. This observation already suggests that there might be
a deeper relation between the chiral dynamics in the matter sector and the confining 
dynamics in the gauge sector and serves as a motivation for the present study.

\begin{figure*}[t]
  \hspace*{0.0cm}
\includegraphics[width=0.41\linewidth]{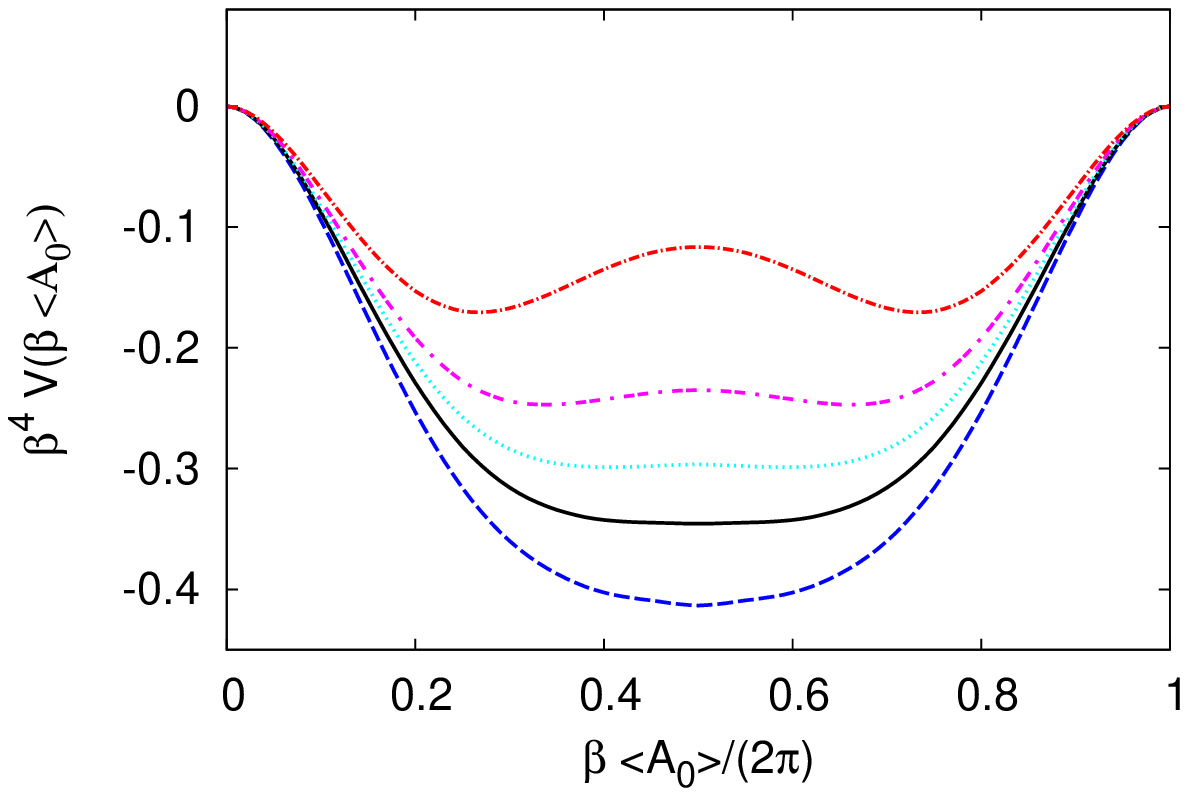}
  \hspace*{2.5cm}
\includegraphics[width=0.41\linewidth]{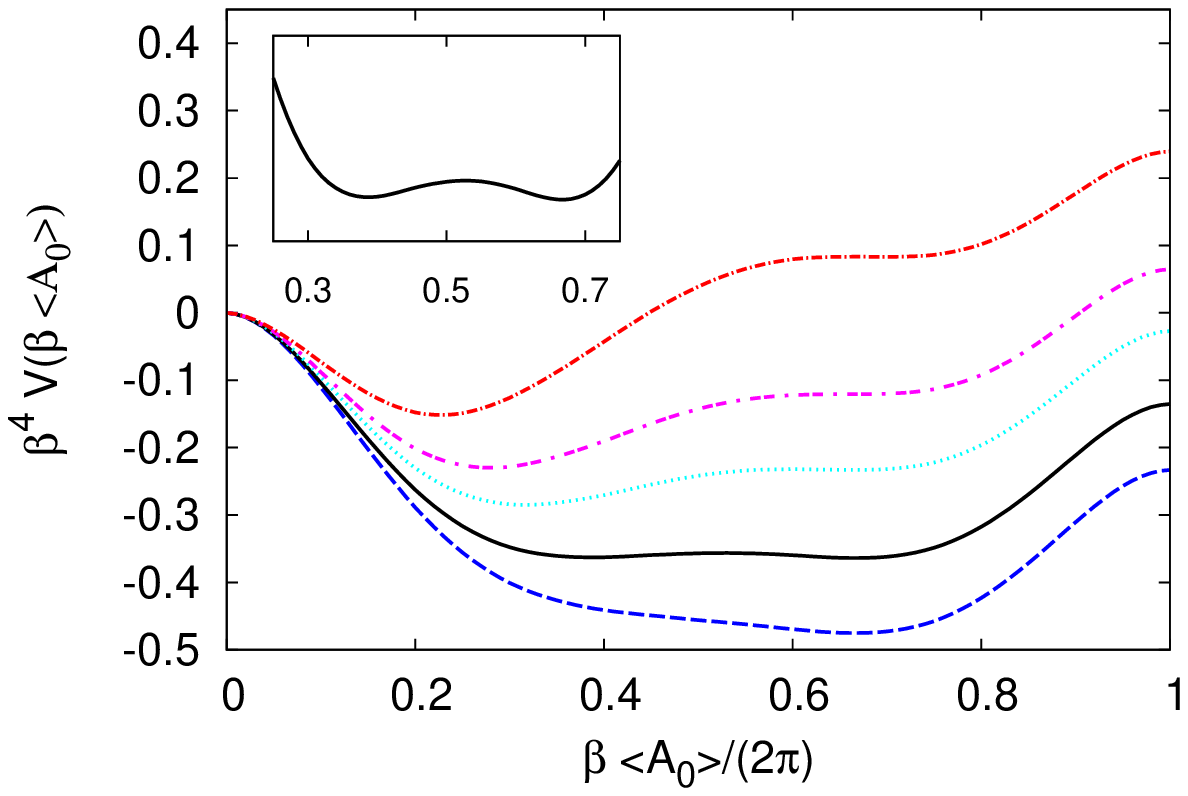} 
\caption{ Normalized order-parameter potential $V(\beta\langle A_0\rangle)$ for $SU(2)$ (left panel) and SU(3)
  (right panel) Yang-Mills theories for various temperatures as obtained
from a first-principles RG study~\cite{Braun:2007bx}. For SU(2) we show the
  potential for temperatures between $T=260\,\text{MeV}$ and $T=285\,\text{MeV}$ (from
  bottom to top). For $SU(3)$ the Cartan subalgebra is two-dimensional and, in turn, 
the potential depends on two independent variables. Here, we show only a slice of
the potential for various temperatures between $T=285\,\text{MeV}$ and $T= 310\,\text{MeV}$
(from bottom to top) in the relevant direction of the Cartan subalgebra. The first-order
phase transition in $SU(3)$ Yang-Mills theory is indicated by a jump in the position 
of the minimum, see also the inlay of the figure. The phase transition temperature is
$\Td\approx 266\,\text{MeV}$ for $SU(2)$ and $\Td\approx 290\,\text{MeV}$ for 
$SU(3)$, respectively.
} 
\label{fig:potentials}
\end{figure*}

The deconfinement phase transition has been studied in pure $SU(\Nc)$ gauge theories and
in QCD with lattice simulations, see 
e.~g. Refs.~\cite{Cheng:2006qk,Aoki:2006br,Aoki:2006we,Aoki:2009sc,Panero:2009tv,Cheng:2009zi,Datta:2010sq,Borsanyi:2010zi,Bazavov:2010pg,Kanaya:2010qd,Bornyakov:2011yb}, 
as well as functional continuum methods~\cite{Braun:2007bx,Marhauser:2008fz,Fischer:2009wc,Fischer:2009gk,Braun:2009gm,Fischer:2010fx,Braun:2010cy,Pawlowski:2010ht}.
In so-called PNJL and Polyakov-loop extended quark-meson (QM) \mbox{models} a 
background field~$\langle A_0\rangle$ is introduced
to study some aspects of quark confinement and the associated phase transition. 
This background field can be related to the so-called Polyakov variable $L[A_0]$,
\be
L[A_0]=\frac{1}{\Nc}\, {\mathcal P}\,\exp\left({\rm i}\bar{g}\int_0^{\beta}
dx_0\,A_0(x_0,\vec{x})
\right)\,,
\ee
where $\beta$ is the inverse temperature, $\Nc$ is the number of colors, 
$\bar{g}$ denotes the bare gauge coupling and $\mathcal P$ stands for path ordering.
In fact, it has been shown that $\tr_{\rm F}\,L[\bfe]$ serves as an order parameter 
for quark confinement in Polyakov-Landau-DeWitt gauge~\cite{Braun:2007bx,Marhauser:2008fz} 
where $\bfe$ is an element of the Cartan subalgebra and 
denotes the ground state of the associated order-parameter potential in the adjoint gauge 
algebra.\footnote{Strictly speaking, we have to distinguish between the background temporal gauge field 
in Landau-DeWitt gauge and its expectation value associated with the order parameter for confinement, 
see Refs.~\cite{Braun:2007bx,Braun:2010cy}. We skip this subtlety here since it is of 
no importance for the present paper and refer to $\bfe$ as the position of the ground-state 
of the order-parameter potential.} This potential can be computed, e.~g., from the knowledge of gauge correlation functions.
In Fig.~\ref{fig:potentials} we show the results for the order-parameter potential as obtained
from a first-principles RG study~\cite{Braun:2007bx,Braun:2010cy}.
The order parameter $\tr_{\rm F}\,L[\bfe]$ is related to the standard 
Polyakov loop $\langle \tr_{\rm F}\,L[A_0]\rangle$ via the Jensen inequality,
\be
\tr_{\rm F}\,L[\bfe]\geq \langle \tr_{\rm F}\, L[A_0]\rangle\,.
\ee

In PNJL/PQM model studies one of the underlying approximations is to set
$\tr_{\rm F}\,L[\bfe] = \langle \tr_{\rm F}\,L[A_0]\rangle$. 
This opens up the possibility to incorporate results for 
the Polyakov loop $\langle \trf L[A_0]\rangle$ as obtained from lattice simulations in these
studies~\cite{Meisinger:1995ih,Pisarski:2000eq,Mocsy:2003qw,Fukushima:2003fw,Megias:2004hj,%
Ratti:2005jh,Sasaki:2006ww,Schaefer:2007pw,Mizher:2010zb,Skokov:2010wb,Herbst:2010rf,Skokov:2010uh}. 
It is then found that the chiral and the deconfinement phase transition lie indeed 
close to each other at small values of the quark chemical potential, as one would naively expect
it to be the case in full QCD~\cite{Cheng:2006qk,Aoki:2006br,Aoki:2006we,Aoki:2009sc,Cheng:2009zi,Borsanyi:2010zi,Bazavov:2010pg,Kanaya:2010qd,Bornyakov:2011yb}. 
However, the inclusion of the Polyakov loop in PNJL/PQM model studies by means of a Polyakov-loop potential is not unique and the QCD phase boundary
at finite chemical potential has indeed been found to be very sensitive to different parameterizations of this
potential, see e.~g. Ref.~\cite{Schaefer:2009ui}. On the other hand the data for the order 
parameter $L[\bfe]$ is available. Therefore it seems natural 
to study at least some of the consequences of the 
approximation~$\tr_{\rm F}\,L[\bfe] = \langle \tr_{\rm F}\,L[A_0]\rangle$
underlying these model studies.

Recently, so-called dual observables arising from a variation of the 
boundary conditions of the fermions in time-like direction have been 
introduced~\cite{Gattringer:2006ci} and employed for a study of the relation 
of quark confinement and chiral symmetry 
breaking at finite temperature~\cite{Synatschke:2007bz,Bilgici:2008qy,Kashiwa:2008bq,Sakai:2008py,Bilgici:2009tx,%
Braun:2009gm,Fischer:2009wc,Fischer:2009gk,Fischer:2010fx,Zhang:2010ui,Mukherjee:2010cp,Gatto:2010qs}. 
These dual observables relate the spectrum of the 
Dirac operator to the order parameter for confinement, namely the Polyakov loop. 
The introduction of these observables constitutes an important formal advance
which allows us to gain a deeper insight into the underlying dynamics at the QCD phase boundary. 
However, they do not allow us to fully resolve the question regarding the relation of 
quark confinement and chiral symmetry breaking. 

In this work we aim to shed more light on the question under which circumstances
the chiral and the deconfinement transition lie close to each other\footnote{In
full QCD with light but finite quark masses both transitions are crossovers.
Since there is no unique way to define the critical temperature associated with a crossover,
a proof of an exact coincidence of the two transitions seems to be impossible in any case.}.
To this end, we analyze the deformation of the RG fixed-point structure of 
chiral four-fermion interactions due to confining gauge dynamics in Sect.~\ref{sec:fermFP}.
As detailed in Refs.~\cite{Gies:2002hq,Gies:2005as,Braun:2005uj,Braun:2006jd,Braun:2008pi,Kondo:2010ts} 
these four-fermion interactions can then be easily connected to the QCD Lagrangian
at high momentum scales. 
In Sect.~\ref{sec:HSTFP} we then present our results
for a partially bosonized formulation of the ansatz discussed in Sect.~\ref{sec:fermFP}. 
In particular, a necessary condition for an exact mapping of both the fermionic
and the partially bosonized theory is discussed.
As one of our main results, we present a phase diagram spanned by the temperature and
the pion decay constant~$f_{\pi}$. The latter is directly related to the chiral condensate.
This phase diagram allows us to gain some insight into the interrelation of quark confinement
and chiral symmetry breaking. In contrast to the phenomenologically more relevant phase diagram 
spanned by the temperature and quark chemical potential, the $(T,f_{\pi})$~phase diagram can be studied
by various different approaches without suffering from problems, e.~g., arising from a
complex-valued Dirac operator. A study of the $(T,f_{\pi})$~phase diagram may therefore be
helpful to, e.~g., benchmark continuum approaches with the aid of lattice simulations.
Our concluding remarks and possible future 
extensions of the present study are given in Sect.~\ref{sec:conc}.

\section{Fermionic Fixed-Point Structure and the locking mechanism}\label{sec:fermFP}
We start our discussion of a dynamical locking mechanism for the
chiral phase transition with an analysis of the fixed-point
structure in the matter sector of QCD. For our study we employ an RG
equation for the quantum effective action, the Wetterich 
equation~\cite{Wetterich:1992yh}. The effective action $\Gamma$ then
depends on the RG scale $k$ (infrared cutoff scale) which determines the RG 'time' 
$t=\ln(k/\Lambda)$ with $\Lambda$ being a UV cutoff scale. For reviews on
and introductions to this functional RG approach we refer the reader to
Refs.~\cite{Litim:1998nf,Bagnuls:2000ae,Berges:2000ew,Polonyi:2001se,Delamotte:2003dw,%
Pawlowski:2005xe,Gies:2006wv,Delamotte:2007pf,Rosten:2010vm}.

For our more general discussion in this section, it suffices to consider
the following ansatz for the effective action:
\be
\Gamma_k [\bar{\psi},\psi,\langle A_0\rangle]&=&\int d^4 x \Big\{ Z_{\psi}\bar{\psi}\left(
{\rm i}\partial\fslash + \bar{g}\gamma_0 \langle A_0\rangle\right)
\psi \nn\\
&& \quad + \frac{\bar{\lambda}_{\psi}}{2}\left[ (\bar{\psi}\psi)^2 
- (\bar{\psi}\vec{\tau} \gamma_5\psi)^2\right]\Big\}\,,
\label{eq:fermionic_action}
\ee
For $\bar{\lambda}_{\psi}\equiv 0$ this ansatz can be considered as the microscopic 
matter part of the QCD action functional in Landau-DeWitt gauge. 
In the present work we restrict ourselves to $\Nf=2$ massless quark flavors and $\Nc$ colors.
The $\tau_i$'s represent the Pauli matrices and couple the spinors in flavor space. The coupling $\bar{\lambda}_{\psi}$ is considered
to be RG-scale dependent. Note that fermionic self-interactions are fluctuation-induced in full QCD, e.~g. by two-gluon exchange, 
and are therefore not fundamental, see Refs.~\cite{Gies:2002hq,Gies:2005as,Braun:2005uj,Braun:2006jd,Braun:2008pi,Braun:2009ns} 
for a detailed discussion.  

Our ansatz for the effective action can be considered as the leading order in a 
systematic derivative expansion of the fermionic sector of QCD. 
The associated small parameter of such an expansion is the anomalous dimension $\eta_{\psi}=-\partial_t \ln Z_{\psi}$
of the fermion fields which has indeed been found to be small in earlier studies with vanishing
background field $\langle A_0\rangle$, see Refs.~\cite{Berges:1997eu,Gies:2002hq,Braun:2008pi,Braun:2009si}.
In the following we therefore set the wave-function renormalization $Z_{\psi}\equiv 1$.

In this section we also drop a possible momentum dependence of the four-fermion 
coupling, $\lambda_{\psi}(|p|\ll k)$. This approximation does not permit a study of properties, such as the meson 
mass spectrum, in the chirally broken regime; for example, mesons manifest themselves as momentum singularities 
in the four-fermion couplings. However, the point-like limit can still be a reasonable approximation in the 
chirally symmetric regime above the chiral phase transition which allows us to gain some insight into the question how
the theory approaches the regime with broken chiral symmetry in the ground state~\cite{Braun:2005uj,Braun:2006jd,Braun:2008pi}. 
For our more quantitative analysis in Sect.~\ref{sec:HSTFP} we shall partly resolve the momentum dependence of 
the fermionic interactions in order to gain access to low-energy observables.
At this point it is important to stress that in the point-like limit 
the RG flow of the four-fermion coupling, which signals the onset of chiral symmetry breaking, is completely decoupled
from the RG flow of fermionic $n$-point functions of higher order. For example, $8$-fermion interactions do not contribute
to the RG flow of the coupling $\bar{\lambda}_{\psi}$ in this limit. 

The constant background field $\bfe$ in our ansatz~\eqref{eq:fermionic_action} is an element of 
the Cartan subalgebra and can be parametrized in terms of the corresponding generators 
\be
\beta \bar{g} \bfe &=& 2\pi\sum_{T^a \in {\rm Cartan}} T^a \phi^{(a)}\nn\\
&=& 2\pi\sum_{T^a \in {\rm Cartan}} T^a v^{(a)} |\phi|\,,\quad v^2=1\,,\label{eq:A0span}
\ee
where the $T_a$'s denote the generators of 
the underlying $SU(\Nc)$ gauge group in the fundamental representation\footnote{The dimension 
of the Cartan subalgebra is $N_c-1$.}. It is convenient to introduce the eigenvalues $\nu _l$ of the 
hermitian matrix in~Eq.~\eqref{eq:A0span}:
\be
\nu_l = {\rm spec}\left\{ (T^{a}v^{a})_{ij}\,\; |\; v^2=1 \right\}\,.
\ee
The presence of a finite background field $\bfe$ now yields
a fermion propagator $(\Gamma ^{(2)})^{-1}$ (inverse two-point function) 
which is no longer proportional to the identity $\mathbbm{1}$ in color space. 
However, it can be spanned by the Cartan subalgebra as follows:
\be
&&\hspace*{-0.7cm}\left(\Gamma ^{(2)}[\{\nu_l |\phi|\}]\right)^{-1}_{ij}
=\frac{1}{N_c} \left(\Gamma ^{(2)}_{0}[\{\nu_l |\phi|\}]\right)^{-1} \mathbbm{1}_{ij} \nn\\
&& \qquad\qquad\qquad + \sum_{T^a \in {\rm Cartan}} \left(\Gamma ^{(2)}_{a}[\{\nu_l |\phi|\}]\right)^{-1} T^a_{ij}\,,
\label{eq:prop_exp}
\ee 
Here, the $T^{a}_{ij}$'s denote the generators in the fundamental representation.
The expansion coefficients on the right-hand side of Eq.~\eqref{eq:prop_exp}
can be obtained straightforwardly by using ${\rm tr}_{\rm F}T^{a}T{^b}=\frac{1}{2}\delta_{ab}$
and ${\rm \tr}_{\rm F}T^a =0$. Of course, it is expected from a physical point of view that 
the expansion~\eqref{eq:prop_exp} of the fermion propagator is convenient once
a background field~$\bfe$ is introduced into the theory since the latter distinguishes a direction
in color space. For $\bfe\equiv 0$ we have $(\Gamma ^{(2)}_{a})^{-1}\equiv 0$.

\begin{figure}[t]
\begin{center}
\begin{picture}(0,80)(0,0)
\put(-50,-5){\includegraphics[scale=1]{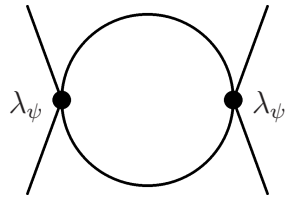}} 
\put(-57,30){  \large$\lambda_{\psi}$}
\put(35,30){  \large$\lambda_{\psi}$}
\end{picture}
\end{center}
\caption{Representation of the $1$PI Feynman diagram associated with the 
$\lambda_{\psi}^2$-term on the right-hand side of the RG flow
  equation~\eqref{eq:lpsi_flow}. Our
    functional RG study includes resummations of this diagram to arbitrary order in~$\lambda_{\psi}$.}
\label{fig:feynman}
\end{figure}

Using the ansatz~\eqref{eq:fermionic_action} together with the parametrization~\eqref{eq:prop_exp},
we obtain the RG flow equation for the dimensionless renormalized four-fermion coupling $\lambda_{\psi}$
in the point-like limit:
\be
&&\hspace*{-0.5cm}\beta_{\lambda_{\psi}}\equiv\partial_t \lambda_{\psi} = (2+2\eta_{\psi})\lambda_{\psi} \nn\\
&& \qquad\qquad\;\; - \frac{2}{\pi ^2}\Big(2 + \frac{1}{\Nc}\Big)
\sum_{l=1}^{N_c} l_{1}^{\rm (F)}(\tau,0,\nu_l |\phi|)\,\lambda_{\psi}^2\,,
\label{eq:lpsi_flow}
\ee
where 
\be
\lambda_{\psi}=Z_{\psi}^{-2} k^2 \bar{\lambda}_{\psi}\,.
\ee
Note that $\lambda_{\psi}$ depends on the background field $\bfe$ and the dimensionless 
temperature $\tau=T/k$. The so-called threshold function $l_{1}^{\rm (F)}$ corresponds
to a one-particle irreducible (1PI) Feynman diagram, see Fig.~\ref{fig:feynman},
and describes the decoupling of massive and thermal modes. 
Moreover, the regularization
scheme dependence is encoded in these functions. 
The definition of the threshold function~$l_{1}^{\rm (F)}$ can be found in App.~\ref{app:thresfcts}.

\begin{figure}[t]
\includegraphics[width=0.9\linewidth]{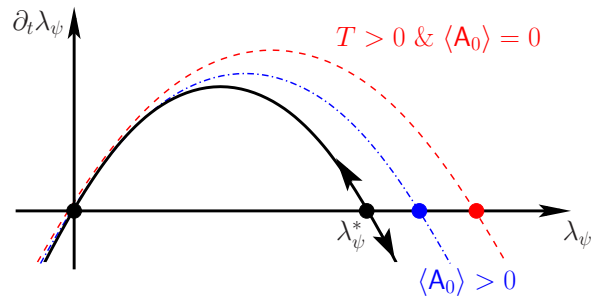}
\caption{Sketch of the $\beta$ function of the four-fermion interaction for 
three different cases: vanishing temperature  
(black/solid line), a given finite value of the temperature $T$ and $\bfe=0$ 
(red/dashed line), the same temperature $T$ but $\bfe >0$ (blue/dashed-dotted line). 
The arrows indicate the direction of the RG flow towards the infrared.}
\label{fig:parabola}
\end{figure}

Let us now discuss the fixed-point structure of the coupling $\lambda_{\psi}$. Apart from a Gau\ss ian 
fixed point we have a second non-trivial fixed point. The
value of this fixed point depends on the dimensionless temperature $\tau$ and the dimensionless 
coordinates $\{\nu_l |\phi|\}$ of the background field $\bfe$, see Fig.~\ref{fig:parabola}.
At vanishing temperature (and background field $\bfe$) we find
\be
\lambda_{\psi}^{\ast}&=&\frac{\pi^2}{(2\Nc +1) l_1^{F}(0,0,0)} + {\mathcal O}(\eta_{\psi}^{\ast})\nn\\
&=&\frac{6\pi^2}{(2\Nc+1)} + {\mathcal O}(\eta_{\psi}^{\ast})
\ee
for the non-Gau\ss ian fixed point. For illustration we have evaluated the threshold function
$l_1^{\rm (F)}$ in the second line for the regulator function~\eqref{eq:fermreg}.
Here, $\eta_{\psi}^{\ast}$ denotes the value of the fermionic anomalous dimension at the fixed point. 
Note that the rescaled fixed-point coupling $\Nc\lfp$ approaches a constant value in the  
limit $\Nc\to\infty$.

The fixed-point value $\lfp$
is not a universal quantity as its dependence on
the threshold function indicates. However, the statement about the existence of the 
fixed point is universal. Choosing an initial value $\luv <\lfp$  
at the initial UV scale $\Lambda$ we find that the theory becomes non-interacting in the
infrared regime ($\lambda_{\psi}\to 0$ for $k\to 0$), see Fig.~\ref{fig:parabola}. For $\lambda_{\psi}^{\rm UV}>\lambda_{\psi}^{\ast}$ 
we find that the four-fermion coupling $\lambda_{\psi}$ increases rapidly and diverges eventually 
at a finite scale $k_{\rm cr}$. This behavior indicates the onset of chiral symmetry breaking
associated with the formation of a quark condensate. Hence chiral symmetry breaking in the IR 
only occurs if we choose $\lambda_{\psi}^{\rm UV}>\lambda_{\psi}^{\ast}$. Of course, the divergence of the 
four-fermion coupling at a finite scale $k_{\rm cr}$ is an artifact of our point-like approximation.
It can be resolved by taking into account (some of) the momentum dependence of the coupling $\lambda_{\psi}$,
see Sect.~\ref{sec:HSTFP}. This will then allow us to gain access to QCD low-energy observables. 
In any case, the scale $k_{\rm cr}$ at which $1/\lambda_{\psi}(k_{\rm cr})=0$ 
sets the scale for a given IR observable~$\mathcal O$:
\be
{\mathcal O}\sim k_{\rm cr}^{d_{\mathcal O}}\,,
\ee
where $d_{\mathcal O}$ is the canonical mass dimension of the observable $\mathcal O$. At vanishing temperature
the scale $k_{\rm cr}$ can be computed analytically. We find
\be
k_{\rm cr}=\Lambda \theta(\lambda_{\psi}^{\rm UV} -\lambda_{\psi}^{\ast})
\left(\frac{\lambda_{\psi}^{\rm UV} -\lambda_{\psi}^{\ast}}{\lambda_{\psi}^{\rm UV}} \right)^{\frac{1}{2}}
+{\mathcal O}(\eta_{\psi}^{\ast})\,.
\label{eq:lambdacr}
\ee
Thus, the critical value $k_{\rm cr}$ scales with the distance of the initial value 
$\lambda_{\psi}^{\rm UV}$ from the fixed-point value $\lfp$. For increasing $\lambda_{\psi}^{\rm UV}$
the scale $k_{\rm cr}$ increases and, in turn, the values of low-energy observables, such 
as the pion decay constant $f_{\pi}$ and the chiral phase transition temperature~$T_{\chi}$, increase.

From now on, we assume that we fix $\lambda_{\psi}^{\rm UV}>\lambda_{\psi}^{\ast}$ at $T=0$. 
The value of $\luv$ then determines the scale $k_{\rm cr}\equiv k_{\rm cr}(\lambda_{\psi}^{\rm UV})$ 
which is related to the values of the low-energy values. For a study of the effects of a finite temperature and 
a finite background field $\bfe$ we then leave our choice for $\luv$ unchanged. This
ensures comparability of the results at zero and finite temperature for a given theory defined by the
choice for $\lambda_{\psi}^{\rm UV}$ at zero temperature.

Next, we turn to a discussion of the fixed-point structure at finite temperature but vanishing
background field~$\bfe$. We still have a Gau\ss ian fixed point. Moreover, we find 
a pseudo fixed-point $\lambda_{\psi}^{\ast}(\tau)$ for arbitrary values of $\tau$ 
at which the right-hand side of the flow equation is zero:
\be
\lambda_{\psi}^{\ast}(\tau) =\frac{\pi^2}{(2\Nc\! +\! 1) l_1^{\rm (F)}(\tau,0,0)  } 
+ {\mathcal O}(\eta_{\psi}^{\ast})\,.
\ee
For high temperatures $T\gg k$ we find $\lambda_{\psi}^{\ast} \sim (T/k)^3$. 
Let us now assume that we have chosen $\luv>\lfp(\tau\!=\! 0)$.
Since the the value of the (pseudo) fixed-point increases with increasing $T/k$, the rapid 
increase of the four-fermion coupling towards the IR ($k\to 0$) is effectively slowed down and may even change its 
direction on the plane defined by the two-dimensional $\beta_{\lambda_{\psi}}$ function\footnote{At finite temperature 
the $\beta_{\lambda_{\psi}}$ function depends on two variables, namely $\tau=T/k$ and $\lambda_{\psi}$.}, 
see also Fig.~\ref{fig:parabola}.
This behavior of the pseudo fixed-point $\lambda_{\psi}^{\ast}(\tau)$ already 
suggests that for a fixed initial value $\luv$ a critical temperature $T_{\chi}$ 
exists above which the four-fermion coupling does not diverge but approaches zero in the IR. 
Such a behavior is indeed expected for high temperatures since the quarks become effectively 
stiff degrees of freedom due to their thermal mass $\sim T$ and chiral symmetry is restored.

Let us now turn to a discussion of the fixed-point structure for finite $T$ and $\bfe$.
In the present approach we consider the value of the background field $\bfe$ as an external input
which is determined by the ground state of the corresponding order-parameter
potential, see also Fig.~\ref{fig:potentials}. As discussed above, the position $\bfe$ of the ground-state 
is directly related to our order parameter for confinement, $\trf L[\bfe]$. 
For temperatures much larger than the deconfinement phase-transition 
temperature~$\Td$ we have $\bfe=0$, i.~e. $\trf L[\bfe] =1$. 
On the other hand, the position $\bfe$ of the ground state in the confined phase of pure $SU(\Nc)$
Yang-Mills theory is uniquely determined up to center transformations by~\cite{Braun:2007bx,Braun:2010cy}
\be
\trf (L[\bfe])^n = 0 
\label{eq:polcoord}
\ee
with $(n\mod\Nc)=1,\dots,\Nc-1$. These conditions determine the $\Nc -1$
coordinates~$\{\phi^{(a)}\}$ of $\bfe$, see Eq.~\eqref{eq:A0span}. 
Moreover, we have
\be
\frac{1}{\Nc}\left|\trf (L[\bfe])^n\right| 
 \leq \frac{1}{\Nc^n}\,. 
\ee
for $n\in {\mathbb N}$. In QCD with dynamical quarks the ground-state 
value~$\bfe$ is then shifted and yields a small but finite order 
parameter in the confined phase. 

For temperatures much larger than the deconfinement phase-transition temperature $\Td$, the fixed-point
structure remains unchanged since $\bfe$ tends to zero for $T\gg \Td$. For finite $\bfe$ 
the pseudo fixed-point $\lambda_{\psi}^{\ast}$ depends on $\bfe$ and $\tau=T/k$. 
The position of the pseudo-fixed point~$\lambda_{\psi}^{\ast}$ can again be given in closed form:
\begin{widetext}
\be
\lambda_{\psi}^{\ast}(\tau,\bfe) &=&\left( \frac{1}{ \pi^2}
\left(2\! +\! \frac{1}{\Nc}\right)\sum_{l=1}^{\Nc} l_1^{\rm (F)}(\tau,0,\nu_l|\phi|)  
\right)^{-1}
\nn\\
&=&\left( \frac{1}{\lambda_{\psi}^{\ast}(0,0)}
+ \frac{1}{6\pi^2}\left(2\!+\!\frac{1}{\Nc}\right) \sum_{n=1}^{\infty}(-\Nc)^{n}\left[\trf (L[\bfe])^n
+ \trf (L^{\dagger}[\bfe])^n\right]
\left(
1+\frac{n}{\tau}
\right){\rm e}^{-\frac{n}{\tau}}
\right)^{-1}\,,
\label{eq:lfpa0}
\ee
\end{widetext}
where we have dropped terms depending on $\eta_{\psi}^{\ast}$ on the right-hand side.
To obtain the second line we have employed the regulator function~\eqref{eq:fermreg}.
However, the general form of the asymptotic series~\eqref{eq:lfpa0} holds for any
regulator function as can be shown by means of Poisson resummation 
techniques\footnote{The sum over the $\tau$-dependent terms in the second line of Eq.~\eqref{eq:lfpa0}
is also closely related to the geometric series.}.
Note that the series~\eqref{eq:lfpa0} can be considered as an expansion for small $\tau=T/k$.

Using Eq.~\eqref{eq:polcoord} we observe that all finite-temperature corrections
to the fixed-point value vanish identically in the confined phase for $\Nc\to\infty$, 
provided that the ground-state value $\bfe$ is identical in $SU(\Nc)$ Yang-Mills theory and
QCD with dynamical fermions. Of course, the latter assumption is not exactly fulfilled
but for physical quark masses it is reasonable to assume 
\be
\trf L[\bfe]\ll 1
\ee
at low and intermediate temperatures. Thus, we have found that 
\be
\lambda_{\psi}^{\ast}(0,0)\equiv\lambda_{\psi}^{\ast}(\tau,\bfe)
\label{eq:fpequiv}
\ee
in the limit $\Nc\to\infty$, independent of the temperature $T$ for $T\lesssim \Td$. 
On the other hand, we have 
\be
\lambda_{\psi}^{\ast}(\tau,\bfe)\to \lambda_{\psi}^{\ast}(\tau,0)\quad\text{for}\quad\bfe\to 0\,.
\ee
With the same reasoning we find
\be
\beta_{\lambda_{\psi}}(0,0)\equiv\beta_{\lambda_{\psi}}(\tau,\bfe)\label{eq:betaequal}
\ee
for $T\lesssim \Td$ and $\Nc\to\infty$. This means that for $T<\Td$ the question of 
whether chiral symmetry is spontaneously broken or not is in fact {\it independent} 
of the temperature, but depends only on the choice of the initial value $\luv$
relative to its fixed-point value~$\lfp$ at $T=0$ and \mbox{$\bfe =0$}. 
We emphasize that Eqs.~\eqref{eq:fpequiv}-\eqref{eq:betaequal} are regularization-scheme independent statements.
For any admissible so-called $3d$ regulator function, e.~g. Eq.~\eqref{eq:fermreg}, 
it is also possible to show analytically that 
\be
\lambda_{\psi}^{\ast}(0,0)\leq\lambda_{\psi}^{\ast}(\tau,\bfe)\leq \lambda_{\psi}^{\ast}(\tau,0)\,
\label{eq:fpineq}
\ee
for arbitrary temperatures $\tau=T/k$ for $\Nc=2,3$; see also discussion of~Eq.~\eqref{eq:l0_FerLoopN} 
in App.~\ref{app:thresfcts}. 
In any case, $\lambda_{\psi}^{\ast}(\tau,\bfe)$
interpolates continuously for a given finite value of $\tau$  between $\lambda_{\psi}^{\ast}(0,0)$ 
and $\lambda_{\psi}^{\ast}(\tau,0)$.

Provided that we choose an initial value $\luv> \lfp(0,0)$, it follows immediately from 
Eq.~\eqref{eq:betaequal} that 
\be
\Tc \geq \Td\,\label{eq:TcTd}
\ee
for $\Nc\to\infty$, see also Ref.~\cite{Meisinger:1995ih}. 
This means that the chiral phase transition is locked in
due to the confining dynamics in the gauge sector.
Loosely speaking, thermal fluctuations of the quark fields, which tend to restore chiral symmetry, are 
suppressed since they are directly linked to the deconfinement order parameter.
Thus, we have found that the restoration of chiral symmetry is intimately connected 
to the confining dynamics in the gauge sector.

In the present analysis the actual chiral phase transition temperature depends on two parameters,
namely the value of the background field $\bfe$ and the initial condition~$\luv$.
However, Eq.~\eqref{eq:TcTd} is a parameter-free statement. It simply follows from an analysis
of the effect of gauge dynamics on the fixed-point structure in the fermionic sector. In particular, we 
have only made use of general properties of the deconfinement order parameter and the fact that
$\luv>\lfp(0,0)$ is a necessary condition for chiral symmetry breaking
at $T=0$ and $\bfe=0$. Of course, the initial condition $\luv$
is not a free parameter in QCD but originally generated by quark-gluon interactions at high (momentum) scales. In a given
regularization scheme the value of $\lambda_{\psi}^{\rm UV}$ can therefore in principle be related to the value of the 
strong coupling $\alpha_{\rm s}$ at, e.~g., the $\tau$ mass scale~\cite{Gies:2005as,Braun:2005uj,Braun:2006jd,Braun:2008pi}.
We would like to point out that neither the value of $\alpha_{\rm s}$ at some scale nor the value of
$\lambda_{\psi}^{\rm UV}$ on a given RG trajectory is a physical observable. However, their values can be related to 
physical low-energy observables. Recall that the value of $\lambda_{\psi}^{\rm UV}$
determines the critical scale $k_{\rm cr}$ which sets the scale for IR observables, see Eq.~\eqref{eq:lambdacr}.

Our findings in the limit $\Nc\to\infty$ even allow us to estimate a window in parameter space
in which the chiral phase transition and the deconfinement phase transition lie close to each other.
In our discussion of chiral symmetry breaking for vanishing background field $\bfe$ we have argued that 
$\Tc\sim k_{\rm cr}$, where the scale $k_{\rm cr}$ is eventually determined by our choice
for $\lambda_{\psi}^{\rm UV}$. If we take into account the
background field~$\bfe$, then the chiral phase transition temperature~$\Tc$ is locked in and
we necessarily have $\Tc\geq \Td$ in the large-$\Nc$ limit, see Eq.~\eqref{eq:TcTd}. Thus, 
the chiral phase transition temperature for all theories which would allow for $\Tc \leq \Td$ for $\bfe=0$ is 
shifted such that $\Tc\simeq\Td$. The upper end of the locking window can therefore
be estimated by the smallest value for $\luv$ for which $\Tc$ for vanishing $\bfe$ is still
larger than~$\Td$. Whereas $\lfp$ and $\luv$ are scheme-dependent quantities, 
the mere existence of such a window in parameter space is a universal statement. 
As we have argued above, the value $\luv$ can be related to the values of physical low-energy
observables. Therefore the existence of such a window for the initial condition $\luv$
suggests the existence of a corresponding window for the values of low-energy observables. We shall come
back to this below when we discuss the phase diagram in the plane spanned by the temperature
and the pion decay constant. 

Let us now discuss the relation of quark confinement and chiral symmetry breaking for finite $\Nc$.
In this case, terms with
\be
n\mod\Nc=0
\ee
contribute to the right-hand side
of Eq.~\eqref{eq:lfpa0} and to the RG flow of $\lambda_{\psi}$. 
We then find that the inequality~\eqref{eq:fpineq} holds only for $\tau=T/k\ll 1$ but not 
for arbitrary values of~$\tau$. 
However, this does not necessarily imply that we do not have a finite range of values for the
initial condition~$\luv$ anymore in which the chiral and the deconfinement phase transition are tightly linked.
It only implies that the lower end of the window for $\luv$ is shifted to larger values
compared to the large-$\Nc$ limit. 

\begin{figure}[t]
  \hspace*{0.0cm}
\includegraphics[width=0.825\linewidth]{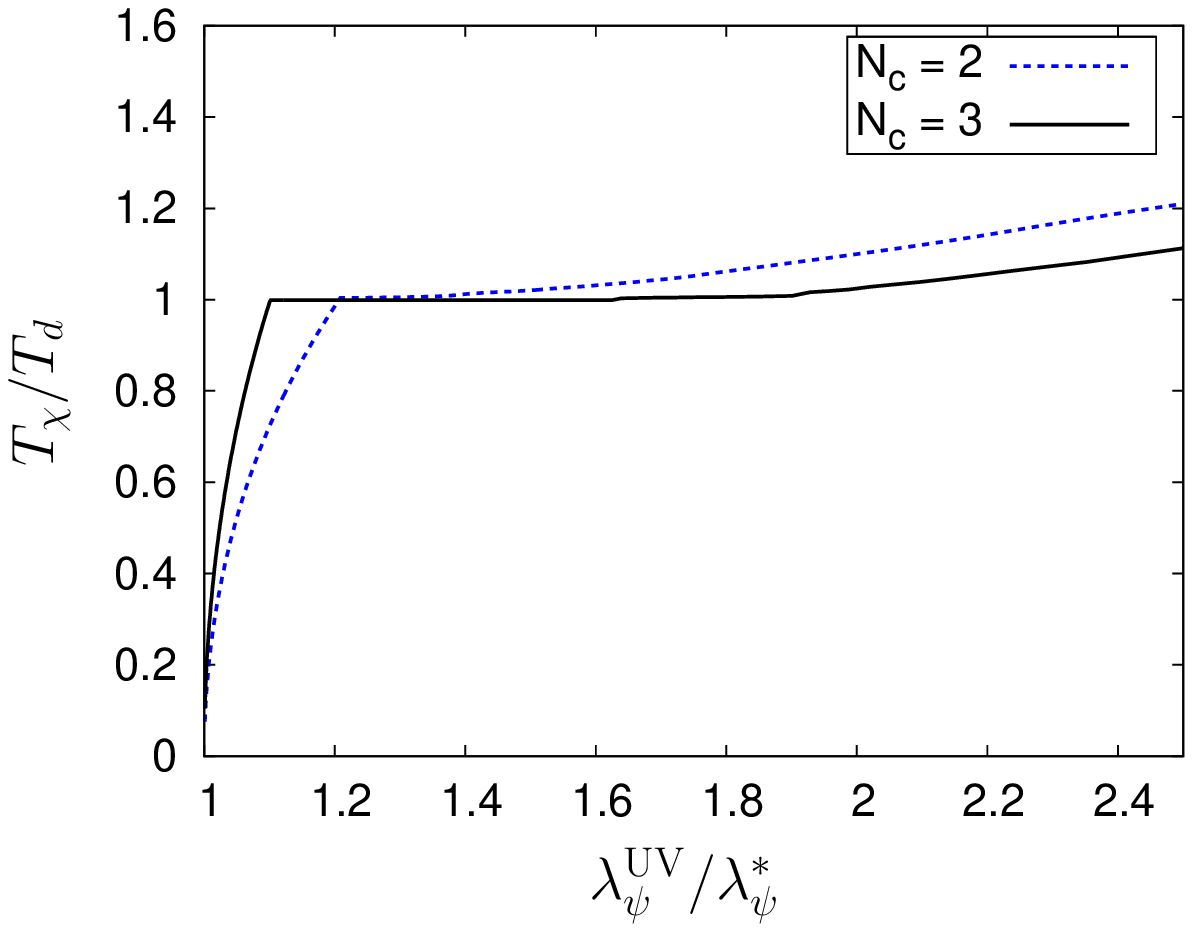}
\caption{Phase diagram for two massless quark flavors 
and $\Nc=2$ colors (blue/dashed line) as well as for $\Nc=3$
colors (black/solid line) in the plane spanned by the
temperature and the rescaled coupling $\luv/\lfp$. The 
lines depict our results for the ratio $\Tc/\Td$ of the chiral and the 
deconfinement phase transition temperature as a function
of $\luv/\lfp$.
Recall that there is no splitting of the phase boundary (i.~e. $\Tc\simeq \Td$) 
for small $\luv$ in the limit~$\Nc\to\infty$.
}
\label{fig:pdlambda}
\end{figure}

To illustrate our analytic findings we have studied numerically the
RG flow of the four-fermion coupling $\lambda_{\psi}$ for finite~$\Nc$. In our setup,
the phase transition temperature is defined to be the smallest temperature
for which~$\lambda_{\psi}$ remains finite in the infrared limit $k\to 0$. Strictly
speaking, this only yields an upper bound for the phase transition temperature
since it is only sensitive to an emergence of a condensate on intermediate 
momentum scales but insensitive to a fate of the condensate in the deep IR due to
fluctuations of the Goldstone modes~\cite{Braun:2009si}.
In Fig.~\ref{fig:pdlambda} we present the phase diagram for two massless quark
flavors and $\Nc=2$ as well as $\Nc=3$ in the plane spanned by the temperature 
and the UV coupling $\luv$. To estimate the phase boundary 
we have employed the results for $\bfe$ for the corresponding $SU(\Nc)$ Yang-Mills theory 
as obtained in Refs.~\cite{Braun:2007bx,Braun:2010cy}, see also Fig.~\ref{fig:potentials}. 
The associated deconfinement phase transition temperature is 
$\Td\approx 266\,\text{MeV}$ for pure gauge $SU(2)$ and $\Td\approx 290\,\text{MeV}$ for 
pure gauge $SU(3)$, respectively. For the UV cutoff we have chosen $\Lambda=1\,\text{GeV}$.
In accordance with our analytic findings we observe that there is only a chirally
symmetric phase for $\luv<\lfp$. 
Taking into account corrections beyond the large-$\Nc$ approximation, we find 
that the chiral phase transition temperature increases continuously with $\luv$, 
$\Tc\! \sim\! (\luv\!-\!\lfp)^{1/2}$, starting from $\Tc=0$ for $\luv=\lfp$. In this regime we have $\Td > \Tc$.
Increasing $\luv$ further we find a window for the values of $\luv$ in which we have $\Tc\sim\Td$. 
In this regime the chiral phase transition is locked in due to the confining dynamics in the gauge sector.
Note that~$\Tc$ is a strictly monotonously increasing function in $\luv$ for $\bfe=0$.
In agreement with our analytic results we observe that the size of this
locking window increases with increasing~$\Nc$. 
For $\luv\gg \lfp$ we then observe that the phase transition temperatures $\Td$ and $\Tc$ differ again.
However, we now have~$\Tc>\Td$.

At this point a few comments are in order. 
First, it is clear that the potential of the confinement order parameter in full QCD (and hence the 
position of the ground-state $\bfe$) also receives contributions from Feynman diagrams
with at least one internal fermion line. These contributions tend to lower the 
deconfinement phase transition temperature\footnote{Strictly speaking we only have
only a deconfinement crossover in the presence of light quarks.}. 
We stress that we only use the Yang-Mills approximation for $\bfe$ 
here to explore the impact of the discussed locking mechanism 
for the chiral phase transition on the finite-temperature phase structure.
In any case, our analytic findings are not (strongly) affected by this approximation
since they rely on very general properties of the confinement order parameter. 
Therefore we still expect that a window in parameter space exists in which the chiral and the deconfinement 
phase transition lie close to each other.
However, our estimate of the phase diagram in the $(T,\luv)$-plane 
and the size of the locking window will change quantitatively when we take into account
the corrections to $\bfe$ due to quark fluctuations. In particular, we expect
that below the locking window the difference~$\Tc$ and~$\Td$ 
becomes smaller since this regime is associated with quarks with a small dynamical 
mass. Second, our ansatz~\eqref{eq:fermionic_action} for the effective action
is not complete with respect to Fierz transformations, see e.~g. Refs.~\cite{Gies:2003dp,Gies:2005as,Braun:2005uj,Braun:2006jd}; 
for example, we have dropped the so-called vector-channel interaction $\sim (\bar{\psi}\gamma_{\mu}\psi)^2$. 
Such interactions would also contribute to the RG flow of the four-fermion interaction $\lambda_{\psi}$. 
At finite temperature the minimal basis of point-like four-fermion interactions is larger than at 
vanishing temperature since the Poincare invariance is broken by the heat bath. 
If we allow for a finite $\bfe$, the minimal set of point-like four-fermion interactions 
is even larger than in the case of vanishing background field $\bfe$. This is due to the
fact that a finite background field $\bfe$ distinguishes a direction in color space. For example, 
our expansion~\eqref{eq:prop_exp} of the fermion propagator suggests that a finite background
field $\bfe$ gives rise to additional point-like interactions of the type~$\sim (\bar{\psi}T^{(3)}\psi)^2$
and~$\sim (\bar{\psi}T^{(8)}\psi)^2$ for $\Nc=3$. However, the additional diagrams are of the same
topology as the one shown in Fig.~\ref{fig:feynman}. We therefore expect that the inclusion of additional 
four-fermion interactions  associated with a Fierz-complete basis is particularly important
for a quantitative computation of the chiral phase transition temperature beyond the large-$\Nc$ limit.
Such an analysis is beyond the scope of this work and deferred to future studies.

Finally we would like to briefly comment on PNJL/PQM-type model studies which are closely
related to the present study. In contrast to the present work, the PNJL/PQM-type
model studies use the assumptions $\trf L[\bfe]=\langle \trf L[A_0]\rangle$ and
$\Nc^n \trf (L[\bfe])^{n}=\Nc\langle \trf L[A_0]\rangle^n$, see e.~g. 
Ref.~\cite{Meisinger:1995ih}. If we used these assumptions, we would 
not have observed the regime with $\Tc < \Td$ for small values of~$(\luv\!-\!\lfp)$ and finite $\Nc$ where
the two phase transitions are decoupled. As we have discussed above, the decoupling of the
phase transitions for small values of~$(\luv\!-\!\lfp)$ is only absent in the limit $\Nc\to\infty$ 
in our study. Therefore we conclude that these approximations used in PNJL/PQM-type model
studies correspond to a large-$\Nc$ approximation in the coupling of the matter and
the gauge sector. This type of large-$\Nc$ approximation should not be confused with the 
large-$\Nc$ approximations used in the matter sector of these models, such as neglecting 
pion fluctuations. In any case, our analysis shows that the large-$\Nc$ approximation associated with the
approximation~$\trf L[\bfe]=\langle \trf L[A_0]\rangle$ clearly affects the dynamics
near the finite-temperature phase boundary and may therefore also affect the predictions
for the $(T,\mu)$ phase diagram from PNJL/PQM-type models. 

\section{The Locking Mechanism and the $(T,f_{\pi})$ Phase diagram}\label{sec:HSTFP}
In the previous section we have discussed the interrelation of quark confinement
and chiral symmetry breaking in a purely fermionic language. We have found that there exists 
a regime in the phase diagram spanned by the temperature and the coupling $\luv$ in which
the chiral and the deconfinement phase transition are tightly linked. In this section, we would like
to map this phase diagram onto a phase diagram spanned by the temperature
and an IR observable of QCD, e.~g. the chiral condensate $|\langle \bar{\psi}\psi\rangle|^{1/3}$ or
the pion decay constant $f_{\pi}$. 

In QCD we have only one input parameter, e.~g. the strong coupling $\alpha_{\rm s}$ at a (high) momentum
scale or, \mbox{equivalently}, $\Lambda_{\rm QCD}$. For vanishing current quark masses $\Lambda_{\rm QCD}$ determines
all physical observables, such as the pion decay constant $f_{\pi}\sim\Lambda_{\rm QCD}$ as well as
the deconfinement and the chiral phase transition temperature, $\Td \sim \Lambda_{\rm QCD}$ and $\Tc \sim \Lambda_{\rm QCD}$.
Thus, real QCD in a phase diagram spanned by the temperature~$T/\Lambda_{\rm QCD}$ 
and $f_{\pi}/\Lambda_{\rm QCD}$ is a single point. In the present paper the scale $\Lambda_{\rm QCD}$ is fixed by the
input for the background field~$\bfe$. In contradistinction to real QCD, however, an additional parameter is present in our 
study, namely~$\luv$. In the following we shall consider this parameter as an asset which allows us to deform QCD. In our model, 
two-flavor QCD as defined by $f_{\pi}\approx 90\,\text{MeV}$ 
then corresponds to a specific choice for $\luv$ on a given RG trajectory.
Thus, our deformed model effectively depends on two parameters, namely $\luv$ and $\Lambda_{\rm QCD}$.
In particular, the chiral phase transition temperature depends on two parameters in our study, 
namely $\luv$ and $\Lambda_{\rm QCD}$, whereas the deconfinement phase transition temperature depends 
only on $\Lambda_{\rm QCD}\sim \Td$. In the remainder of the paper we exploit the dependence of our model on $\luv$ in more detail 
to gain some insight into the relation of chiral and confining dynamics close to the finite-temperature phase boundary.
Eventually, this leads to a prediction for a $(T,f_{\pi})$ phase diagram. 
We would like to add that de\-formations of QCD-like \mbox{theories} with 
an additional relevant parameter, such as a four-fermion coupling, indeed play a prominent role
in beyond-standard model applications, see e.~g. Ref.~\cite{Fukano:2010yv}. 

For the computation of the $(T,f_{\pi})$ phase diagram we employ  a (partially) bosonized version of our ansatz~\eqref{eq:fermionic_action}.
Whereas the purely fermionic description is particularly convenient for analytic studies, the 
partially bosonized ansatz allows us to resolve momentum dependences
of fermionic self-interactions in a simple manner and therefore permits a study of the formation of the chiral 
condensate and the mass spectrum of mesons. To obtain the partially bosonized formulation 
of our ansatz~\eqref{eq:fermionic_action} we perform a Hubbard-Stratonovich transformation
of the underlying path-integral~\cite{Hubbard:1959ub,Stratonovich}. 
This introduces auxiliary fields $\bar{\Phi}^{T}=(\sigma,\vec{\pi})$ 
into the theory which mediate the interaction between the quarks. 
Here, we assume that the bosons are composites of fermions and 
do not carry an internal charge, e.~g. color or flavor: $\sigma\sim (\bar{\psi}\psi)$ and 
$\vec{\pi}\sim (\bar{\psi} \vec{\tau}\gamma_5\psi)$. The components of $\bar{\Phi}$ 
are labeled according to the role that the corresponding fields are playing in the spontaneously
broken regime. The effective action of the partially bosonized theory then 
reads\footnote{At finite temperature the Poincare invariance of the theory is broken. Therefore 
the wave-function renormalizations longitudinal and transversal to the heat-bath
obey a different RG running. We neglect this difference in the present work. This is justified since 
it has indeed been found in Ref.~\cite{Braun:2009si} that the difference is small at low temperatures 
and only yields mild corrections to, e.~g., the
thermal mass of the bosonic degrees of freedom for intermediate temperatures~$T\gtrsim\Tc$.}
\be
\Gamma_k[\bar{\psi},\psi,\bar{\Phi},\bfe]
&=& \int d^4 x\,\Big\{ Z_{\psi}\bar{\psi}\left(\I \partial\!\!\!\slash + \bar{g}\gamma_0 \langle A_0\rangle\right)\psi\nn\\
&& +\,\frac{1}{2}Z_{\Phi}\left(\partial_{\mu}\bar{\Phi}\right)^2
 + {\rm i}\bar{h}\bar{\psi}(\sigma\!+\! {\rm i}\vec{\tau}\cdot\vec{\pi}\gamma_5)\psi\nn\\
&& \quad\quad +\, \frac{1}{2}\bar{m}^2\bar{\Phi}^2 + \frac{1}{8}\bar{\lambda}_{\Phi}\bar{\Phi}^4 \Big\}\,,\label{Eq:HSTAction}
\ee
with a Yukawa coupling $\bar{h}\in \mathbb{R}$ and the boundary conditions
\be
\lim_{k\to \Lambda} Z_{\Phi} &=& 0\,,\label{eq:bc1}
\\
\lim_{k\to \Lambda} Z_{\psi} &=& 1\,,\label{eq:bc2}
\\
\lim_{k\to \Lambda} \bar{\lambda}_{\Phi} &=& 0\,.\label{eq:bc3}
\ee
These boundary conditions together with the identity
\be
\bar{\lambda}_{\psi}=\frac{\bar{h}^2}{\bar{m}^2}\label{eq:hstmap}
\ee
allow us to map the ansatz~\eqref{Eq:HSTAction} onto the 
model~\eqref{eq:fermionic_action} at the initial UV scale $\Lambda$. In particular, we are now able to 
relate the initial values $\luv$ to physically meaningful IR observables, e.~g. the pion decay constant
and the chiral condensate.

The term $\sim \bar{\Phi}^4$ in the effective action~\eqref{Eq:HSTAction} corresponds to an $8$-fermion interaction term  
in the purely fermionic description, $\bar{\Phi} \sim \bar{\psi}\psi$. Due to the boundary condition~\eqref{eq:bc3} this term is generated
dynamically and not adjusted by hand in our RG approach, 
see e.~g.~\cite{Berges:1997eu,Gies:2002hq,Braun:2008pi,Braun:2009si,Braun:2009gm}.
Thus, the value of the corresponding coupling at the initial RG scale $\Lambda$ does not represent an additional parameter
of the theory. In any case, this coupling only  plays a prominent role in the deep IR regime where it 
accounts for the mass difference between the $\sigma$-meson and the \mbox{pions}.

A word of caution concerning the mapping of the partially bosonized and the purely fermionic description 
needs to be added here. The identity~\eqref{eq:hstmap} suggests that we indeed have only one input parameter
in the present study, namely the value of the ratio $\bar{h}^2_{\rm UV}/\bar{m}^2_{\rm UV}$ at the UV scale~$\Lambda$.
In $d=4$ space-time dimensions, however, 
the Yukawa coupling $\bar{h}$ is marginal. This suggests that the partially bosonized theory in $d=4$ depends on 
two input parameters in contrast to $d=3$, see Refs.~\cite{Braun:2009si,Braun:2010tt}. 
Nonetheless the critical scale~$k_{\rm cr}$ only depends on the 
ratio~$\bar{h}^2_{\rm UV}/\bar{m}^2_{\rm UV}$ in leading order in an expansion in powers of $1/\Nc$ and receives only small 
corrections from the next-to-leading order, see below. On the other hand, 
the ratio of IR observables, such as the ratio of the $\sigma$~mass and the constituent quark mass, 
depends on both parameters~\cite{Braun:2010tt}. In the following we choose the ratio~$\bar{h}^2_{\rm UV}/\bar{m}^2_{\rm UV}$ and the Yukawa 
coupling~$\bar{h}_{\rm UV}$ as the independent input parameters at the UV scale~$\Lambda$. 

Let us now discuss the RG flow equations of the dimensionless renormalized couplings $m^2=\bar{m}^{2}/(Z_{\Phi}k^2)$, 
$\lambda_{\Phi}=\bar{\lambda}_{\Phi}/Z_{\Phi}^2$ and the Yukawa coupling $h=\bar{h}/(Z_{\Phi}^{1/2}Z_{\psi})$. 
Using the ansatz~\eqref{Eq:HSTAction} together with the parametrization~\eqref{eq:prop_exp} of the fermion
propagator we find
\be
\partial _t m^2 &=& (\eta_{\Phi}-2)m^2 - \frac{3}{2\pi^2} l_1(\tau,m^2) \lambda_{\Phi} \nn\\
&& \quad\quad +\,\frac{4}{\pi ^2}
\sum_{l=1}^{N_c} l_{1}^{\rm (F)}(\tau,m^2_{\rm q},\nu_l |\phi|) h^2\,,\label{eq:m2flow}
\\
\partial _t \lambda_{\Phi}&=& 2 \eta_{\Phi}\lambda_{\Phi} + \frac{3}{\pi^2} l_2(\tau,m^2) \lambda_{\Phi}^2 \nn\\
&& \quad\quad -\,\frac{8}{\pi ^2}
\sum_{l=1}^{N_c} l_{2}^{\rm (F)}(\tau,m^2_{\rm q},\nu_l |\phi|) h^4\,,\label{eq:lflow}
\\
\partial _t h^2 &=& (\eta_{\Phi} +2\eta_{\psi})h^2
 \nn\\
 && \;\; -\,\frac{2}{\pi ^2}\frac{1}{\Nc}\sum_{l=1}^{N_c} 
 l_{1,1}^{\rm  (FB)}(\tau,m^2_{\rm q},\nu_l |\phi|,m^2)  h^4 ,
\label{eq:hflow}
\ee
where $\eta_{\Phi}=-\partial _t \ln Z_{\Phi}$ and $\eta_{\psi}=-\partial _t \ln Z_{\psi}$. The threshold functions
are defined in App.~\ref{app:thresfcts}. Note that the quarks are massless in the symmetric
regime, $m_{\rm q}\equiv 0$.

At this point it is instructive to have a closer look at the mapping between the partially bosonized and the purely fermionic
description. To this end, we consider the RG flow of the ratio $h^2/m^2$ which can be obtained straightforwardly from
the flow equations~\eqref{eq:m2flow} and~\eqref{eq:hflow}. We obtain
\be
&&\partial_t \left(\frac{h^2}{m^2}\right)=(2\!+\! 2\eta_{\psi})\left(\frac{h^2}{m^2}\right)
\!+\! \frac{3}{2\pi^2} l_1(\tau,m^2) \lambda_{\Phi}  \left(\frac{h^2}{m^4}\right) 
 \nn\\ 
 &&\qquad\qquad\qquad\quad
 - \frac{4}{\pi ^2}
\sum_{l=1}^{N_c} l_{1}^{\rm (F)}(\tau,m^2_{\rm q},\nu_l |\phi|)\left(\frac{h^2}{m^2}\right)^2 \nn\\
&& \qquad\quad
-\frac{2}{\pi ^2}\frac{1}{\Nc}\sum_{l=1}^{N_c} 
 l_{1,1}^{\rm  (FB)}(\tau,m^2_{\rm q},\nu_l |\phi|,m^2)  \left(\frac{h^4}{m^2}\right).
\label{eq:h2m2flow}
\ee
Using Eqs.~\eqref{eq:bc3} and~\eqref{eq:hstmap} as well as 
\be
l_{1,1}^{\rm  (FB)}(\tau,m^2_{\rm q},\nu_l |\phi|,m^2) \stackrel{ (m\gg 1)}{\longrightarrow}\frac{1}{m^2}\, l_{1}^{\rm (F)}(\tau,m^2_{\rm q},\nu_l |\phi|)\,,
\ee
we recover the RG flow equation~\eqref{eq:lpsi_flow}
of the four-fermion coupling~$\lambda_{\psi}$. Thus, the partially bosonized and the purely fermionic
description are indeed identical at the UV scale $\Lambda$. Note that the prefactor of the term 
\mbox{$\sim h^2/m^2$} would turn out to be incorrect if we did not include the RG running of the
Yukawa coupling. In other words, a standard local potential approximation does not incorporate
all terms associated with a systematic expansion of the flow equations in powers of~$1/\Nc$. In
Ref.~\cite{Braun:2010tt} this observation is discussed in the context of the Gross-Neveu model.
 
From Eq.~\eqref{eq:h2m2flow} we also deduce that the partially bosonized
description allows us to go conveniently beyond the point-like approximation employed 
in the purely fermionic description discussed in Sect.~\ref{sec:fermFP}. 
To be more specific, we observe that the momentum dependence of the four-fermion vertex is effectively parametrized
by the RG flow of the four-boson coupling $\lambda_{\Phi}$, the Yukawa coupling~$h$ 
and the mass parameter~$m^2$. We stress that the purely fermionic point-like description and 
the partially bosonized description are no longer identical for scales $k<\Lambda$. From the 
flow equation of the ratio $h^2/m^2$, however, it follows that
the differences are quantitatively small in the symmetric regime since the renormalized mass
of the mesons is large.

In the regime with broken chiral symmetry in the ground state the mass parameter $m^2$ assumes 
negative values. This behavior signals the existence of a finite
vacuum expectation value of the field $\Phi$. 
In this regime, it is therefore convenient to study the RG flow of the four-boson coupling $\lambda_{\Phi}$
and the vacuum expectation value $\langle\Phi\rangle\equiv\Phi_0$. The RG flow of the latter
can be obtained from the following condition:
\be
\frac{d}{dt}\left[\frac{\partial}{\partial\bar{\Phi}^2}\left(
\frac{1}{2}\bar{m}^2\bar{\Phi}^2 + \frac{1}{8}\bar{\lambda}_{\Phi}\bar{\Phi}^4
\right)\right]_{\bar{\Phi}_0}\stackrel{!}{=}0\,.
\ee
The flow equations for $\Phi_0$ and $\lambda_{\Phi}$ are then given by
\be
&&\partial _t \Phi_0^2=-(\eta_{\Phi}+2)\Phi_0^2 + \frac{3}{2\pi^2}  l_1(\tau,m_{\sigma}^2)
+  \frac{3}{2\pi^2}  l_1(\tau,m_{\pi}^2) \nn\\
&& \qquad\qquad\qquad\qquad - \frac{8}{\pi ^2}
\sum_{l=1}^{N_c} l_{1}^{\rm (F)}(\tau,m_{\rm q}^2,\nu_l |\phi|) \frac{h^2}{\lambda_{\Phi}}\,,
\ee
\be
&&\partial _t \lambda_{\Phi}= 2 \eta_{\Phi}\lambda_{\Phi} + \frac{9}{4\pi^2} l_2(\tau,m_{\sigma}^2) \lambda_{\Phi}^2
 + \frac{3}{4\pi^2} l_2(\tau,m_{\pi}^2) \lambda_{\Phi}^2 \nn\\
&& \qquad\qquad\qquad\qquad -\,\frac{8}{\pi ^2}
\sum_{l=1}^{N_c} l_{2}^{\rm (F)}(\tau,m_{\rm q}^2,\nu_l |\phi|) h^4\,,
\ee
where $m_{\sigma}^2=\lambda_{\Phi}\Phi_0^2$, $m_{\pi}^2=0$ and
$m_{\rm q}^2=h^2\Phi_0^2$ is the constituent quark mass. Note that
we identify $Z_{\Phi}^{-1/2}\bar{\Phi}_0$ with the pion decay constant~$f_{\pi}$. For simplicity, we
neglect the running of the Yukawa coupling in the regime with a broken chiral symmetry of the
ground-state. In fact, $1$PI diagrams with at least one internal quark line are
parametrically suppressed in this regime since the quarks acquire a (large) mass. Therefore the RG flow
of the theory in the spontaneously broken regime is mainly governed 
by $1$PI diagrams with internal pion lines only. However, the
latter class of diagrams does not directly contribute to the RG flow of the Yukawa 
coupling\footnote{In the symmetric regime we have only taken into account the contributions
to the running of the Yukawa coupling~$h$ which are required to map the flow of the 
ratio~$h^2/m^2$ onto the RG flow of the four-fermion coupling $\lambda_{\psi}$ at the 
UV scale~$\Lambda$.}. 

\begin{figure}[t]
  \hspace*{0.0cm}
\includegraphics[width=0.825\linewidth]{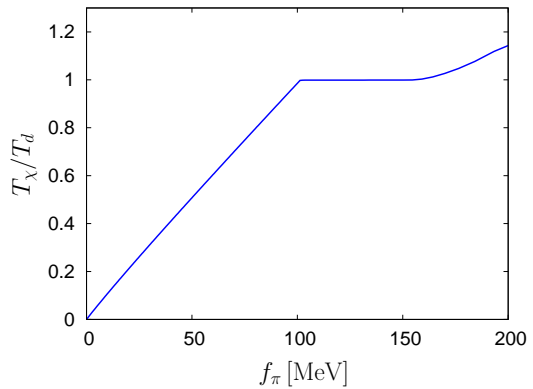}
\caption{Phase diagram for two massless quark flavors 
and $\Nc=3$ in the plane spanned by the rescaled 
temperature $\Tc/\Td$ and the value of the pion decay constant $f_{\pi}$
at $T=0$. The line depicts our result for the ratio $\Tc/\Td$ of the chiral and
the deconfinement phase transition temperature.
}
\label{fig:pdhst}
\end{figure}

In the present study we also neglect the running of the bosonic and fermionic wave-function 
renormalizations. Thus, we consider $Z_{\Phi}\equiv 1$ and $Z_{\psi}\equiv 1$.  
In studies with vanishing gluonic background field~\cite{Berges:1997eu,Gies:2002hq,Braun:2008pi}
it has indeed been found that the associated anomalous dimensions $\eta_{\Phi}$ and $\eta_{\psi}$ are small 
over a wide range of scales. Nonetheless using $Z_{\Phi}\equiv 1$ 
means that we violate the initial condition $m^2 \gg 1$ for the renormalized mass at the UV scale~$\Lambda$.
The latter condition is equivalent to the condition~\eqref{eq:bc1}. Therefore our study of the partially bosonized
theory necessarily relies on two parameters instead of one, e.~g. the initial values for $h^2$ and $h^2/m^2$,
which we expect to be anyway the case, see discussion above.
Employing $Z_{\Phi}\equiv 1$ then seems to be a reasonable
approximation for a first explorative study of the locking mechanism discussed in Sect.~\ref{sec:fermFP}.
In contrast to the flow of $Z_{\Phi}$, the RG flow of 
$Z_{\psi}$ is solely driven by $1$PI diagrams
with at least one internal boson and one internal fermion line.
Since the mesons are heavy in the symmetric regime and
the fermions are heavy in the regime with broken chiral symmetry in the ground state,
it is justified to consider $Z_{\psi}\equiv 1$ for our purposes. For a more quantitative study
including a determination of critical exponents, however, the running of $Z_{\psi}$, $Z_{\Phi}$
and the Yukawa coupling needs to be taken into account in both regimes since very close to
the phase transition both the mesons and the quarks are approximately massless. The corresponding
RG flow equations can be derived along the lines of, e.~g., Refs.~\cite{Berges:1997eu,Gies:2002hq,Braun:2008pi}.

Using the partially bosonized theory we now estimate the size of the locking window for the chiral 
phase transition in terms of the pion decay constant $f_{\pi}$.
The latter is directly related to the chiral condensate $|\langle \bar{\psi}\psi\rangle|^{1/3}$ via the
Gell-Mann-Oakes-Renner relation. In Fig.~\ref{fig:pdhst} we present our result for the phase boundary for
two massless quark flavors and $\Nc=3$ in the plane spanned by the temperature and the value of the
pion decay constant at $T=0$. This phase diagram corresponds
to the phase diagram discussed in Sect.~\ref{sec:fermFP}. To obtain Fig.~\ref{fig:pdhst} we have set $\Lambda=1\,\text{GeV}$
and used $h_{\rm UV}=3$ as initial condition at $k=\Lambda$. Different initial values 
for $\lambda_{\psi}^{\rm UV}=h^2_{\rm UV}/m^2_{\rm UV}$ then translate into different values for~$f_{\pi}$.
In addition, we have employed the results for~$\bfe$ as obtained from a RG study of $SU(3)$ Yang-Mills 
theory~\cite{Braun:2007bx,Braun:2010cy} which is in very good agreement with lattice QCD results.
We neglect the back-reaction of the matter sector on $\bfe$ and use only the Yang-Mills 
approximation for $\bfe$ here to explore the impact of the discussed locking mechanism on the finite-temperature phase structure.

In Fig.~\ref{fig:pdhst} we notice that the chiral phase transition temperature tends to 
zero for $f_{\pi}\to 0$ and that $\Tc$ is a monotonic function for small values of $f_{\pi}$, $\Tc\sim f_{\pi}$.
For  $0\leq f_{\pi}\lesssim 100\,\text{MeV}$ and  $f_{\pi}\gtrsim 150\,\text{MeV}$ we observe a splitting
of the phase boundary. Interestingly, we have $\Tc > \Td$ 
for~$f_{\pi}\gtrsim 150\,\text{MeV}$ and $\Tc < \Td$ for $f_{\pi}\lesssim 100\,\text{MeV}$.
As a non-trivial result 
we find that the chiral phase transition is locked to the deconfinement phase transition
for $100\,\text{MeV} \lesssim f_{\pi}\lesssim 150\,\text{MeV}$.\footnote{Using $h_{\rm UV}=3.5$ as initial condition for the Yukawa coupling,
we find that the value of the lower end of this window for $f_{\pi}$ is lowered by approximately $10\%$.
However, the size of this window is roughly independent of the considered initial conditions for the Yukawa
coupling.}  This observation confirms
our results from the study of the fixed-point structure in the purely fermionic description. Moreover, we 
find that the physically relevant range of values of the pion decay constant lies right below the locking window. 
The near coincidence of the two phase transition temperatures for $f_{\pi}\sim 90\,\text{MeV}$ has also been
found in an RG study including the back-reaction of the quark fluctuations on the order-parameter potential~\cite{Braun:2009gm}.
Recall that for $f_{\pi}\lesssim 150\,\text{MeV}$ the chiral phase transition temperature is 
shifted to larger values compared to studies with $\bfe\equiv 0$ since the thermal excitation of 
quarks is suppressed by the confining dynamics in the gauge sector, see Eq.~\eqref{eq:lfpa0}. Loosely speaking,
we observe a competition between the confining dynamics and the thermal excitation of the system. While an increase
of the temperature tends to restore chiral symmetry, the confining dynamics in the gauge sector favors a ground
state with broken chiral symmetry.

For $f_{\pi}\gtrsim 150\,\text{MeV}$, which corresponds to constituent quark masses $m_{\rm q}\gtrsim 450\,\text{MeV}$, 
we find a second-order chiral phase transition for vanishing current quark masses. 
Moreover, our result for $\Tc$ as a function of $f_{\pi}$ does not depend
on the presence of the gluonic background field~$\bfe$ any more. 
In this regime, we expect that our results are least affected by our approximation of neglecting the back-reaction 
of the matter sector on the deconfinement order parameter. Here, the dynamics in the gauge sector and the matter sector 
are expected to be effectively decoupled. 
Nonetheless the difference between~$\Tc$ and~$\Td$ might be smaller in a full QCD study.

For $f_{\pi}\lesssim 150\,\text{MeV}$ a window may open up in which the chiral phase transition is 
of first order. This is due to the significant shift of the chiral phase transition temperature to larger values 
in this regime compared to the case without a gluonic background field. Provided the deconfinement order parameter rises
quickly above the (deconfinement) phase transition~\cite{Braun:2007bx,Marhauser:2008fz,Braun:2010cy,Dumitru:2010mj}, 
strong thermal excitations may induce a chiral transition of first order. 
In our present study we indeed find a first-order chiral phase-transition for $100\,\text{MeV} \lesssim f_{\pi}\lesssim 150\,\text{MeV}$.
However, we rush to add that this observation is a shortcoming of our approximations since we have employed the results for~$\bfe$
as obtained from a study of $SU(3)$ Yang-Mills theory but neglected the back-reaction of the matter sector
on~$\bfe$. Thus, the first-order phase transition in the gauge sector induces a first-order chiral phase transition in this regime.
Nevertheless a first-order chiral phase transition may occur in this regime even if we include the
back-reaction of the matter sector on the confinement order parameter, provided the latter rises rapidly for $T\gtrsim\Td$.
For example, this might be the case for physical pion masses. In this respect the present analysis
provides a simple mechanism for a first-order chiral phase transition~\cite{D'Elia:2005bv}.
For $f_{\pi}\lesssim 100\,\text{MeV}$ we find again a second-order chiral phase transition for 
vanishing current quark masses.

We would like to remind the reader of the fact that we have only one parameter in QCD with
massless current quarks and that four-fermion interactions are induced by fluctuations, e.~g. two-gluon exchange;
see Refs.~\cite{Gies:2002hq,Gies:2005as,Braun:2005uj,Braun:2006jd,Braun:2008pi,Braun:2009ns} for a detailed discussion.
In the present paper we have considered the initial value $\luv$ of the four-fermion coupling as an
additional parameter which allowed us to deform QCD and study some aspects of the relation of
quark confinement and chiral symmetry breaking. In any case, we would like to stress that the relation of the 
fixed-point structure in the matter sector to the deconfinement order parameter is a universal statement which 
is {\it a priori} independent of the details of the actual scale fixing procedure. This interplay of the matter 
and the gauge sector suggests the existence of a window for the values of the physical 
observables in which the chiral and the deconfinement phase transition lie close to each other.

\section{Conclusions and Outlook}\label{sec:conc}
In the present paper we have analyzed the interplay of the deconfinement and the chiral phase transition. 
To this end, we have studied the fixed-point structure of four-fermion interactions 
in the presence of a finite temporal gluonic background field. The latter can be directly related to an order
parameter for the deconfinement phase transition, namely $\trf L[\bfe] \geq \langle \trf L[A_0]\rangle$.
In a purely fermionic description the onset of chiral symmetry breaking is indicated by rapidly increasing 
four-fermion interactions.  Restoration of chiral symmetry is then displayed by four-fermion couplings which approach a 
Gau\ss ian fixed point in the IR. Thus, the question of chiral symmetry breaking in the IR can ultimately be 
linked to  the fixed-point structure of the four-fermion couplings.

We have indeed found that the fixed-point structure of four-fermion interactions is directly related to the 
order parameter for confinement. In particular, we have analyzed the scalar-pseudoscalar interaction
channel and have shown that its $\beta$~function does not depend on the temperature in the limit $\Nc\to\infty$,
provided we are in the confined phase, i.~e. $\trf L[\bfe]\to 0$. Loosely speaking, thermal excitations
of the quarks are suppressed in the confined phase. Therefore the chiral phase transition is locked in
and shifted to higher temperatures. Our findings confirm the results of a mean-field study 
by Meisinger and Ogilvie~\cite{Meisinger:1995ih} in which it has been pointed out that the chiral order-parameter
potential is independent of the temperature in the confined phase by employing the assumption~$\trf L[\bfe]=\langle \trf L[A_0]\rangle$.
Our non-perturbative analysis of the fixed-point structure of the matter sector establishes this observation.
However, we have not made use of the assumption~$\trf L[\bfe]=\langle \trf L[A_0]\rangle$. In fact, our study shows that 
the latter assumption is only justified in the limit $\Nc\to\infty$. Moreover, we have studied how 
corrections to the large-$\Nc$ approximation affect the dynamics at the finite-temperature phase boundary.
The locking mechanism discussed in Sect.~\ref{sec:fermFP} eventually allows us to determine a  
window for the values of the pion decay constant in which the chiral and the
deconfinement phase transition lie close to each other. The size of this window depends on the number
of colors $\Nc$ and is maximal in the limit~$\Nc\to\infty$. 

For $\Nc=3$ and two massless quark flavors
we have computed the phase diagram in the plane spanned by the temperature and the pion decay constant.
This phase diagram can be divided into three different regimes. For small values of $f_{\pi}$ we find a regime
with $\Td>\Tc$. For $100\,\text{MeV}\lesssim f_{\pi}\lesssim 150\,\text{MeV}$ the deconfinement and the chiral
phase transition lie close to each other. Finally, there is a third regime for $f_{\pi}\gtrsim 150\,\text{MeV}$
which is characterized by $\Tc>\Td$. Here, the chiral condensate is large and therefore the matter and the 
gauge sector are effectively decoupled. 

Of course, the present study can be improved in many 
ways. Currently we are including the contributions to the RG flow of the four-fermion coupling arising 
from $1$PI diagrams with one- or two internal gluon lines~\cite{BraunJanot}. This will eventually allow
us to get rid of the parameter $\luv$ so that we are left with a single parameter for the gauge and the matter
sector, namely $\alpha_{\rm s}$ at a given UV scale.

While a study of the QCD phase diagram in the $(T,\mu)$ plane is of great phenomenological importance,
we think that a study of the phase diagram in the $(T,f_{\pi})$ plane (or equivalently in 
the $(T,|\langle \bar{\psi}\psi\rangle|^{1/3})$ plane) may provide us with important insights concerning
the interplay of chiral and confining dynamics. In particular, such a study of the $(T,f_{\pi})$ phase
diagram does not suffer from problems arising, e.~g., from a complex-valued spectrum of the Dirac operator as
it is the case at finite quark chemical potential~$\mu$. Therefore, an analysis of the $(T,f_{\pi})$ 
phase diagram may be also helpful to benchmark results from continuum 
approaches against those from lattice QCD simulations.

\acknowledgments 
The authors are very grateful to H.~Gies, B.~Klein and J.~M.~Pawlowski 
for useful discussions and a critical reading of the manuscript. 
JB acknowledges support by the DFG research training group GRK 1523/1.

\appendix
\section{Threshold functions}\label{app:thresfcts}

In this appendix we summarize technical details concerning the derivation of the 
RG flow equations. 

In the computation of the RG flow equations a regulator function needs to be specified
which determines the regularization scheme~\cite{Wetterich:1992yh}.
In the present work we have employed an optimized spatial regulator function
for the bosonic as well as for the fermionic degrees of 
freedom~\cite{Litim:2000ci,Litim:2001fd,Litim:2001up,Litim:2006ag,Blaizot:2006rj}. 
For the bosons, we choose
\be
\!\!\!\!\!\!\! R_{\rm B}(\vec{p}^{\,2})=\vec{p}^{\,2}\left(\frac{k^2}{\vec{p}^{\,2}}\!-\!1\right)\theta(k^2\!-\!\vec{p}^{\,2})
\equiv \vec{p}^{\,2} r_{\rm B} \left({ \frac{\vec{p}^{\,2}}{k^2}}\right),
\label{eq:bosreg}
\ee
whereas we choose
\be
\!\!\!\!\!\!\! R_{\psi}(\vec{p})=\vec{p}\fslash\left(\sqrt{\frac{k^2}{\vec{p}^{\,2}}}\!-\!1\right)\theta(k^2\!-\!\vec{p}^{\,2})
\equiv \vec{p}\fslash\, r_{\psi} \left({ \frac{\vec{p}^{\,2}}{k^2}}\right)
\label{eq:fermreg}
\ee
for the fermionic degrees of freedom. In the following we define the threshold functions
relevant for the present work. These functions represent the $1$PI diagrams contributing 
to the RG flow of the studied couplings. For a generalizations of the threshold functions
to an arbitrary number of space-time dimensions we refer the reader to Ref.~\cite{Braun:2008pi}
where also the dependence of these functions on the anomalous dimensions is displayed.
The latter is of no relevance for the present study.

In order to define the threshold functions, it is convenient to define dimensionless propagators for the 
bosons~(B) and the fermions~($\psi$), respectively:
\be
\tilde{G} _{\rm B} (x_0,\omega)=\frac{1}{ x_0 + x(1+r_{\rm B}) + \omega}
\ee 
and
\be
\tilde{G} _{\psi} (x_0,\omega)=\frac{1}{ x_0 + x(1+r_{\psi})^2 + \omega}\,,
\ee
where $x=\vec{p}^{\,2}/k^2$.

First, we define the threshold functions which appear in the RG flow equations for the
bosonic self-interactions. For the purely bosonic loops, we find
\be
l_0 (\tau,\omega)&=&\frac{\tau}{2}\sum_{n=-\infty}^{\infty}\int _0 ^{\infty} dx\, x^{\frac{3}{2}} 
(\partial _t r_{\rm B} )\,\tilde{G} _{\rm B} (\tilde{\omega}_n^2,\omega)\nn\\
&=&\frac{2}{3}\frac{1}{\sqrt{1+\omega}}
\left(\frac{1}{2} + \bar{n}_{\rm B}(\tau,\omega) \right)\label{eq:l0_BosLoop}
\ee
where $\tau=T/k$ denotes the dimensionless temperature and $\tilde{\omega}_n=2\pi n\tau$ 
denotes the dimensionless bosonic Matsubara frequencies. The function $\bar{n}_{\rm B}$ 
represents the Bose-Einstein distribution function:
\be
\bar{n}_{\rm B}(\tau,\omega)=\frac{1}{\E ^{\sqrt{1+\omega}/\tau} -1}\,.
\ee
Bosonic threshold functions of order $n$ are then derived from Eq.~\eqref{eq:l0_BosLoop} by taking
derivatives with respect to the mass parameter $\omega$:
\be
\frac{\partial}{\partial \omega} l_n (\tau,\omega) 
= -(n + \delta_{n,0})\, l_{n+1}  (\tau,\omega)\,.
\ee
For the purely fermionic loops contributing to the flow equations of the bosonic self-interactions but also
to the RG flow of the four-fermion coupling, we find
\be
&& l_0 ^{(\rm F)} (\tau,\omega,\mu)=\tau\sum_{n=-\infty}^{\infty}\int _0 ^{\infty} dx\, x^{\frac{3}{2}} 
(\partial _t r_{\psi} )(1+r_{\psi})\times\nn\\
&& \qquad\qquad\qquad\qquad\qquad\qquad\qquad\times\tilde{G} _{\psi} ((\tilde{\nu}_n+ 2\pi\tau\mu)^2,\omega)\nn\\
&& \quad=\frac{1}{3}\frac{1}{\sqrt{1+\omega}}
\left(1 \!-\! \bar{n}_{\psi} (\tau,{\rm i}\mu,\omega) \!-\! \bar{n}_{\psi} (\tau,-{\rm i}\mu,\omega) \right)\label{eq:l0_FerLoop}.
\ee
Here, we have introduced the dimensionless fermionic Matsubara frequencies $\tilde{\nu}_n=(2n+1)\pi\tau$.
The function $\bar{n}_{\psi}$ represents the Fermi-Dirac distribution function:
\be
\bar{n}_{\psi}(\tau,\mu,\omega)=\frac{1}{\E ^{(\sqrt{1+\omega}/\tau) + 2\pi\mu} +1}\,.
\ee
Higher-order fermionic threshold functions can again be found by taking derivatives with respect to the
mass parameter $\omega$:
\be
\frac{\partial}{\partial \omega} l_n ^{\rm (F)} (\tau,\omega,\mu) 
= -(n + \delta_{n,0})\, l_{n+1}^{\rm (F)} (\tau,\omega,\mu)\,.\label{eq:l0_FerLoopN}
\ee
To prove Eq.~\eqref{eq:fpineq} it is convenient to perform the sum over Matsubara frequencies. The 
threshold function $l_1 ^{\rm (F)}$  can then be written as follows
\be
&&l_1 ^{\rm (F)}(\tau,0,\mu)=\int _0 ^{\infty} dx\, x^{\frac{3}{2}} (\partial _t r_{\psi} )(1\!+\! r_{\psi})\Bigg\{
\frac{1}{4 [f(x)]^{\frac{3}{2}}}\nn\\
&&\qquad\quad\; -\frac{1} {2 \tau [f(x)]^{3/2}  [g(x,\mu)]^2}
\bigg[2 \tau  {\rm e}^{\frac{2\sqrt{f(x)}}{\tau }} [\cos (2 \pi  \mu )]^2\nn\\
&&\qquad\qquad\qquad\quad +\,{\rm e}^{\frac{\sqrt{f(x)}}{\tau }}
   \left(\sqrt{f(x)} \left({\rm e}^{\frac{2 \sqrt{f(x)}}{\tau }}+1\right)\right.\nn\\
 &&\qquad\quad\; \left.\,+\tau  \left({\rm e}^{\frac{2
   \sqrt{f(x)}}{\tau }}+3\right)\right) \cos (2 \pi  \mu )+ \tau  {\rm e}^{\frac{2 \sqrt{f(x)}}{\tau }}\nn\\
 &&\qquad\qquad\qquad\qquad  +\, 2\sqrt{f(x)} {\rm e}^{\frac{2 \sqrt{f(x)}}{\tau }}+\tau\bigg]
   \Bigg\}\,,\label{eq:lf1}
\ee
where $f(x)=x(1\!+\! r_{\psi})^2$ and 
\be
g(x,\mu)=2\,{\rm e}^{-\frac{\sqrt{f(x)}}{\tau }} \cos (2 \pi  \mu )+{\rm e}^{-\frac{2 \sqrt{f(x)}}{\tau }}+1\,.
\ee
The first term in the curly bracket on the right-hand side of Eq.~\eqref{eq:lf1} is the zero-temperature contribution which
is strictly positive. The second term (square bracket) corresponds to the finite-temperature corrections and is strictly negative,
provided that $\cos (2\pi\mu)>0$. For regulator functions with $(\partial_t r_{\psi})\geq 0$ we
then find that the finite-temperature corrections yield a negative contribution to $ l_1 ^{\rm (F)}$. In the RG flow equation of the
four-fermion coupling we sum over the eigenvalues $\mu=\nu_l |\phi|$. Since 
\be
\sum_{l=1} ^{\Nc} \cos(2\pi\nu_l |\phi|) \geq 0
\ee
for $\Nc=2$ and $\Nc=3$, we have proven~Eq.~\eqref{eq:fpineq}.

Finally we give the definition of the threshold function which appears in the RG flow equations of
the Yukawa coupling. We have
\be
&&l_{1,1}^{\rm (FB)}(\tau,\omega_{\psi},\mu,\omega_{\rm B})\nn\\
&&\quad=-\frac{\tau}{2}\,\sum_{n=-\infty}^{\infty}
\int _0 ^{\infty} dx\, x^{\frac{1}{2}}\tilde{\partial}_t \,\Big\{\tilde{G} _{\psi} ((\tilde{\nu}_n+2\pi\tau \mu)^2,\omega_{\psi})\times\nn\\
&& \qquad\qquad\qquad\qquad\qquad\qquad\qquad\times\tilde{G} _{\rm B} (\tilde{\nu}_n^2,\omega_{\rm B})\Big\}\,.
\ee
To evaluate the integral over $x$ (spatial momenta), we have to take derivatives with respect to the regulator 
function. For the regulator functions~\eqref{eq:bosreg} and~\eqref{eq:fermreg} these derivatives are given by 
\be
\tilde{\partial}_t \Big|_{\psi}&=& \frac{1}{x^{1/2}}\theta(1-x)\frac{\partial}{\partial r_{\psi}}\,,\\
\tilde{\partial}_t \Big|_{\rm B}&=&\frac{2}{x} \theta(1-x)\frac{\partial}{\partial r_{\rm B}}\,,
\ee
where the first and the second line defines how the formal derivative $\tilde{\partial} _t$ acts on 
the fermion propagator and the boson propagator, respectively.


\begin{thebibliography}{76}
\expandafter\ifx\csname natexlab\endcsname\relax\def\natexlab#1{#1}\fi
\expandafter\ifx\csname bibnamefont\endcsname\relax
  \def\bibnamefont#1{#1}\fi
\expandafter\ifx\csname bibfnamefont\endcsname\relax
  \def\bibfnamefont#1{#1}\fi
\expandafter\ifx\csname citenamefont\endcsname\relax
  \def\citenamefont#1{#1}\fi
\expandafter\ifx\csname url\endcsname\relax
  \def\url#1{\texttt{#1}}\fi
\expandafter\ifx\csname urlprefix\endcsname\relax\def\urlprefix{URL }\fi
\providecommand{\bibinfo}[2]{#2}
\providecommand{\eprint}[2][]{\url{#2}}

\bibitem[{\citenamefont{Braun-Munzinger
  et~al.}(2003)\citenamefont{Braun-Munzinger, Redlich, and
  Stachel}}]{BraunMunzinger:2003zd}
\bibinfo{author}{\bibfnamefont{P.}~\bibnamefont{Braun-Munzinger}},
  \bibinfo{author}{\bibfnamefont{K.}~\bibnamefont{Redlich}}, \bibnamefont{and}
  \bibinfo{author}{\bibfnamefont{J.}~\bibnamefont{Stachel}}
  (\bibinfo{year}{2003}), \eprint{nucl-th/0304013}.

\bibitem[{\citenamefont{McLerran and Pisarski}(2007)}]{McLerran:2007qj}
\bibinfo{author}{\bibfnamefont{L.}~\bibnamefont{McLerran}} \bibnamefont{and}
  \bibinfo{author}{\bibfnamefont{R.~D.} \bibnamefont{Pisarski}},
  \bibinfo{journal}{Nucl. Phys.} \textbf{\bibinfo{volume}{A796}},
  \bibinfo{pages}{83} (\bibinfo{year}{2007}), \eprint{0706.2191}.

\bibitem[{\citenamefont{Meisinger and Ogilvie}(1996)}]{Meisinger:1995ih}
\bibinfo{author}{\bibfnamefont{P.~N.} \bibnamefont{Meisinger}}
  \bibnamefont{and} \bibinfo{author}{\bibfnamefont{M.~C.}
  \bibnamefont{Ogilvie}}, \bibinfo{journal}{Phys. Lett.}
  \textbf{\bibinfo{volume}{B379}}, \bibinfo{pages}{163} (\bibinfo{year}{1996}),
  \eprint{hep-lat/9512011}.

\bibitem[{\citenamefont{Pisarski}(2000)}]{Pisarski:2000eq}
\bibinfo{author}{\bibfnamefont{R.~D.} \bibnamefont{Pisarski}},
  \bibinfo{journal}{Phys. Rev.} \textbf{\bibinfo{volume}{D62}},
  \bibinfo{pages}{111501} (\bibinfo{year}{2000}), \eprint{hep-ph/0006205}.

\bibitem[{\citenamefont{Mocsy et~al.}(2004)\citenamefont{Mocsy, Sannino, and
  Tuominen}}]{Mocsy:2003qw}
\bibinfo{author}{\bibfnamefont{A.}~\bibnamefont{Mocsy}},
  \bibinfo{author}{\bibfnamefont{F.}~\bibnamefont{Sannino}}, \bibnamefont{and}
  \bibinfo{author}{\bibfnamefont{K.}~\bibnamefont{Tuominen}},
  \bibinfo{journal}{Phys. Rev. Lett.} \textbf{\bibinfo{volume}{92}},
  \bibinfo{pages}{182302} (\bibinfo{year}{2004}), \eprint{hep-ph/0308135}.

\bibitem[{\citenamefont{Fukushima}(2004)}]{Fukushima:2003fw}
\bibinfo{author}{\bibfnamefont{K.}~\bibnamefont{Fukushima}},
  \bibinfo{journal}{Phys. Lett.} \textbf{\bibinfo{volume}{B591}},
  \bibinfo{pages}{277} (\bibinfo{year}{2004}), \eprint{hep-ph/0310121}.

\bibitem[{\citenamefont{Megias et~al.}(2006)\citenamefont{Megias, Ruiz~Arriola,
  and Salcedo}}]{Megias:2004hj}
\bibinfo{author}{\bibfnamefont{E.}~\bibnamefont{Megias}},
  \bibinfo{author}{\bibfnamefont{E.}~\bibnamefont{Ruiz~Arriola}},
  \bibnamefont{and} \bibinfo{author}{\bibfnamefont{L.~L.}
  \bibnamefont{Salcedo}}, \bibinfo{journal}{Phys. Rev.}
  \textbf{\bibinfo{volume}{D74}}, \bibinfo{pages}{065005}
  (\bibinfo{year}{2006}), \eprint{hep-ph/0412308}.

\bibitem[{\citenamefont{Ratti et~al.}(2006)\citenamefont{Ratti, Thaler, and
  Weise}}]{Ratti:2005jh}
\bibinfo{author}{\bibfnamefont{C.}~\bibnamefont{Ratti}},
  \bibinfo{author}{\bibfnamefont{M.~A.} \bibnamefont{Thaler}},
  \bibnamefont{and} \bibinfo{author}{\bibfnamefont{W.}~\bibnamefont{Weise}},
  \bibinfo{journal}{Phys. Rev.} \textbf{\bibinfo{volume}{D73}},
  \bibinfo{pages}{014019} (\bibinfo{year}{2006}), \eprint{hep-ph/0506234}.

\bibitem[{\citenamefont{Sasaki et~al.}(2007)\citenamefont{Sasaki, Friman, and
  Redlich}}]{Sasaki:2006ww}
\bibinfo{author}{\bibfnamefont{C.}~\bibnamefont{Sasaki}},
  \bibinfo{author}{\bibfnamefont{B.}~\bibnamefont{Friman}}, \bibnamefont{and}
  \bibinfo{author}{\bibfnamefont{K.}~\bibnamefont{Redlich}},
  \bibinfo{journal}{Phys. Rev.} \textbf{\bibinfo{volume}{D75}},
  \bibinfo{pages}{074013} (\bibinfo{year}{2007}), \eprint{hep-ph/0611147}.

\bibitem[{\citenamefont{Schaefer et~al.}(2007)\citenamefont{Schaefer,
  Pawlowski, and Wambach}}]{Schaefer:2007pw}
\bibinfo{author}{\bibfnamefont{B.-J.} \bibnamefont{Schaefer}},
  \bibinfo{author}{\bibfnamefont{J.~M.} \bibnamefont{Pawlowski}},
  \bibnamefont{and} \bibinfo{author}{\bibfnamefont{J.}~\bibnamefont{Wambach}},
  \bibinfo{journal}{Phys. Rev.} \textbf{\bibinfo{volume}{D76}},
  \bibinfo{pages}{074023} (\bibinfo{year}{2007}), \eprint{0704.3234}.

\bibitem[{\citenamefont{Mizher et~al.}(2010)\citenamefont{Mizher, Chernodub,
  and Fraga}}]{Mizher:2010zb}
\bibinfo{author}{\bibfnamefont{A.~J.} \bibnamefont{Mizher}},
  \bibinfo{author}{\bibfnamefont{M.~N.} \bibnamefont{Chernodub}},
  \bibnamefont{and} \bibinfo{author}{\bibfnamefont{E.~S.} \bibnamefont{Fraga}},
  \bibinfo{journal}{Phys. Rev.} \textbf{\bibinfo{volume}{D82}},
  \bibinfo{pages}{105016} (\bibinfo{year}{2010}), \eprint{1004.2712}.

\bibitem[{\citenamefont{Skokov et~al.}(2010{\natexlab{a}})\citenamefont{Skokov,
  Stokic, Friman, and Redlich}}]{Skokov:2010wb}
\bibinfo{author}{\bibfnamefont{V.}~\bibnamefont{Skokov}},
  \bibinfo{author}{\bibfnamefont{B.}~\bibnamefont{Stokic}},
  \bibinfo{author}{\bibfnamefont{B.}~\bibnamefont{Friman}}, \bibnamefont{and}
  \bibinfo{author}{\bibfnamefont{K.}~\bibnamefont{Redlich}},
  \bibinfo{journal}{Phys. Rev.} \textbf{\bibinfo{volume}{C82}},
  \bibinfo{pages}{015206} (\bibinfo{year}{2010}{\natexlab{a}}),
  \eprint{1004.2665}.

\bibitem[{\citenamefont{Herbst et~al.}(2011)\citenamefont{Herbst, Pawlowski,
  and Schaefer}}]{Herbst:2010rf}
\bibinfo{author}{\bibfnamefont{T.~K.} \bibnamefont{Herbst}},
  \bibinfo{author}{\bibfnamefont{J.~M.} \bibnamefont{Pawlowski}},
  \bibnamefont{and} \bibinfo{author}{\bibfnamefont{B.-J.}
  \bibnamefont{Schaefer}}, \bibinfo{journal}{Phys. Lett.}
  \textbf{\bibinfo{volume}{B696}}, \bibinfo{pages}{58} (\bibinfo{year}{2011}),
  \eprint{1008.0081}.

\bibitem[{\citenamefont{Skokov et~al.}(2010{\natexlab{b}})\citenamefont{Skokov,
  Friman, and Redlich}}]{Skokov:2010uh}
\bibinfo{author}{\bibfnamefont{V.}~\bibnamefont{Skokov}},
  \bibinfo{author}{\bibfnamefont{B.}~\bibnamefont{Friman}}, \bibnamefont{and}
  \bibinfo{author}{\bibfnamefont{K.}~\bibnamefont{Redlich}}
  (\bibinfo{year}{2010}{\natexlab{b}}), \eprint{1008.4570}.

\bibitem[{\citenamefont{Gies and Wetterich}(2004)}]{Gies:2002hq}
\bibinfo{author}{\bibfnamefont{H.}~\bibnamefont{Gies}} \bibnamefont{and}
  \bibinfo{author}{\bibfnamefont{C.}~\bibnamefont{Wetterich}},
  \bibinfo{journal}{Phys. Rev.} \textbf{\bibinfo{volume}{D69}},
  \bibinfo{pages}{025001} (\bibinfo{year}{2004}), \eprint{hep-th/0209183}.

\bibitem[{\citenamefont{Gies and Jaeckel}(2006)}]{Gies:2005as}
\bibinfo{author}{\bibfnamefont{H.}~\bibnamefont{Gies}} \bibnamefont{and}
  \bibinfo{author}{\bibfnamefont{J.}~\bibnamefont{Jaeckel}},
  \bibinfo{journal}{Eur. Phys. J.} \textbf{\bibinfo{volume}{C46}},
  \bibinfo{pages}{433} (\bibinfo{year}{2006}), \eprint{hep-ph/0507171}.

\bibitem[{\citenamefont{Braun and Gies}(2007)}]{Braun:2005uj}
\bibinfo{author}{\bibfnamefont{J.}~\bibnamefont{Braun}} \bibnamefont{and}
  \bibinfo{author}{\bibfnamefont{H.}~\bibnamefont{Gies}},
  \bibinfo{journal}{Phys. Lett.} \textbf{\bibinfo{volume}{B645}},
  \bibinfo{pages}{53} (\bibinfo{year}{2007}), \eprint{hep-ph/0512085}.

\bibitem[{\citenamefont{Braun and Gies}(2006)}]{Braun:2006jd}
\bibinfo{author}{\bibfnamefont{J.}~\bibnamefont{Braun}} \bibnamefont{and}
  \bibinfo{author}{\bibfnamefont{H.}~\bibnamefont{Gies}},
  \bibinfo{journal}{JHEP} \textbf{\bibinfo{volume}{06}}, \bibinfo{pages}{024}
  (\bibinfo{year}{2006}), \eprint{hep-ph/0602226}.

\bibitem[{\citenamefont{Braun}(2009)}]{Braun:2008pi}
\bibinfo{author}{\bibfnamefont{J.}~\bibnamefont{Braun}}, \bibinfo{journal}{Eur.
  Phys. J.} \textbf{\bibinfo{volume}{C64}}, \bibinfo{pages}{459}
  (\bibinfo{year}{2009}), \eprint{0810.1727}.

\bibitem[{\citenamefont{Braun and Gies}(2010)}]{Braun:2009ns}
\bibinfo{author}{\bibfnamefont{J.}~\bibnamefont{Braun}} \bibnamefont{and}
  \bibinfo{author}{\bibfnamefont{H.}~\bibnamefont{Gies}},
  \bibinfo{journal}{JHEP} \textbf{\bibinfo{volume}{05}}, \bibinfo{pages}{060}
  (\bibinfo{year}{2010}), \eprint{0912.4168}.

\bibitem[{\citenamefont{Braun et~al.}(2010{\natexlab{a}})\citenamefont{Braun,
  Fischer, and Gies}}]{Braun:2010qs}
\bibinfo{author}{\bibfnamefont{J.}~\bibnamefont{Braun}},
  \bibinfo{author}{\bibfnamefont{C.~S.} \bibnamefont{Fischer}},
  \bibnamefont{and} \bibinfo{author}{\bibfnamefont{H.}~\bibnamefont{Gies}}
  (\bibinfo{year}{2010}{\natexlab{a}}), \eprint{1012.4279}.

\bibitem[{\citenamefont{Cheng et~al.}(2006)}]{Cheng:2006qk}
\bibinfo{author}{\bibfnamefont{M.}~\bibnamefont{Cheng}} \bibnamefont{et~al.},
  \bibinfo{journal}{Phys. Rev.} \textbf{\bibinfo{volume}{D74}},
  \bibinfo{pages}{054507} (\bibinfo{year}{2006}), \eprint{hep-lat/0608013}.

\bibitem[{\citenamefont{Aoki et~al.}(2006{\natexlab{a}})\citenamefont{Aoki,
  Fodor, Katz, and Szabo}}]{Aoki:2006br}
\bibinfo{author}{\bibfnamefont{Y.}~\bibnamefont{Aoki}},
  \bibinfo{author}{\bibfnamefont{Z.}~\bibnamefont{Fodor}},
  \bibinfo{author}{\bibfnamefont{S.~D.} \bibnamefont{Katz}}, \bibnamefont{and}
  \bibinfo{author}{\bibfnamefont{K.~K.} \bibnamefont{Szabo}},
  \bibinfo{journal}{Phys. Lett.} \textbf{\bibinfo{volume}{B643}},
  \bibinfo{pages}{46} (\bibinfo{year}{2006}{\natexlab{a}}),
  \eprint{hep-lat/0609068}.

\bibitem[{\citenamefont{Aoki et~al.}(2006{\natexlab{b}})\citenamefont{Aoki,
  Endrodi, Fodor, Katz, and Szabo}}]{Aoki:2006we}
\bibinfo{author}{\bibfnamefont{Y.}~\bibnamefont{Aoki}},
  \bibinfo{author}{\bibfnamefont{G.}~\bibnamefont{Endrodi}},
  \bibinfo{author}{\bibfnamefont{Z.}~\bibnamefont{Fodor}},
  \bibinfo{author}{\bibfnamefont{S.~D.} \bibnamefont{Katz}}, \bibnamefont{and}
  \bibinfo{author}{\bibfnamefont{K.~K.} \bibnamefont{Szabo}},
  \bibinfo{journal}{Nature} \textbf{\bibinfo{volume}{443}},
  \bibinfo{pages}{675} (\bibinfo{year}{2006}{\natexlab{b}}),
  \eprint{hep-lat/0611014}.

\bibitem[{\citenamefont{Aoki et~al.}(2009)}]{Aoki:2009sc}
\bibinfo{author}{\bibfnamefont{Y.}~\bibnamefont{Aoki}} \bibnamefont{et~al.},
  \bibinfo{journal}{JHEP} \textbf{\bibinfo{volume}{06}}, \bibinfo{pages}{088}
  (\bibinfo{year}{2009}), \eprint{0903.4155}.

\bibitem[{\citenamefont{Panero}(2009)}]{Panero:2009tv}
\bibinfo{author}{\bibfnamefont{M.}~\bibnamefont{Panero}},
  \bibinfo{journal}{Phys. Rev. Lett.} \textbf{\bibinfo{volume}{103}},
  \bibinfo{pages}{232001} (\bibinfo{year}{2009}), \eprint{0907.3719}.

\bibitem[{\citenamefont{Cheng et~al.}(2010)}]{Cheng:2009zi}
\bibinfo{author}{\bibfnamefont{M.}~\bibnamefont{Cheng}} \bibnamefont{et~al.},
  \bibinfo{journal}{Phys. Rev.} \textbf{\bibinfo{volume}{D81}},
  \bibinfo{pages}{054504} (\bibinfo{year}{2010}), \eprint{0911.2215}.

\bibitem[{\citenamefont{Datta and Gupta}(2010)}]{Datta:2010sq}
\bibinfo{author}{\bibfnamefont{S.}~\bibnamefont{Datta}} \bibnamefont{and}
  \bibinfo{author}{\bibfnamefont{S.}~\bibnamefont{Gupta}},
  \bibinfo{journal}{Phys. Rev.} \textbf{\bibinfo{volume}{D82}},
  \bibinfo{pages}{114505} (\bibinfo{year}{2010}), \eprint{1006.0938}.

\bibitem[{\citenamefont{Borsanyi et~al.}(2010)}]{Borsanyi:2010zi}
\bibinfo{author}{\bibfnamefont{S.}~\bibnamefont{Borsanyi}} \bibnamefont{et~al.}
  (\bibinfo{year}{2010}), \eprint{1011.4230}.

\bibitem[{\citenamefont{Bazavov and Petreczky}(2010)}]{Bazavov:2010pg}
\bibinfo{author}{\bibfnamefont{A.}~\bibnamefont{Bazavov}} \bibnamefont{and}
  \bibinfo{author}{\bibfnamefont{P.}~\bibnamefont{Petreczky}},
  \bibinfo{journal}{PoS} \textbf{\bibinfo{volume}{LATTICE2010}},
  \bibinfo{pages}{169} (\bibinfo{year}{2010}), \eprint{1012.1257}.

\bibitem[{\citenamefont{Kanaya}(2010)}]{Kanaya:2010qd}
\bibinfo{author}{\bibfnamefont{K.}~\bibnamefont{Kanaya}}
  (\bibinfo{year}{2010}), \eprint{1012.4235}.

\bibitem[{\citenamefont{Bornyakov et~al.}(2011)}]{Bornyakov:2011yb}
\bibinfo{author}{\bibfnamefont{V.~G.} \bibnamefont{Bornyakov}}
  \bibnamefont{et~al.} (\bibinfo{year}{2011}), \eprint{1102.4461}.

\bibitem[{\citenamefont{Braun et~al.}(2010{\natexlab{b}})\citenamefont{Braun,
  Gies, and Pawlowski}}]{Braun:2007bx}
\bibinfo{author}{\bibfnamefont{J.}~\bibnamefont{Braun}},
  \bibinfo{author}{\bibfnamefont{H.}~\bibnamefont{Gies}}, \bibnamefont{and}
  \bibinfo{author}{\bibfnamefont{J.~M.} \bibnamefont{Pawlowski}},
  \bibinfo{journal}{Phys. Lett.} \textbf{\bibinfo{volume}{B684}},
  \bibinfo{pages}{262} (\bibinfo{year}{2010}{\natexlab{b}}),
  \eprint{0708.2413}.

\bibitem[{\citenamefont{Marhauser and Pawlowski}(2008)}]{Marhauser:2008fz}
\bibinfo{author}{\bibfnamefont{F.}~\bibnamefont{Marhauser}} \bibnamefont{and}
  \bibinfo{author}{\bibfnamefont{J.~M.} \bibnamefont{Pawlowski}}
  (\bibinfo{year}{2008}), \eprint{0812.1144}.

\bibitem[{\citenamefont{Fischer}(2009)}]{Fischer:2009wc}
\bibinfo{author}{\bibfnamefont{C.~S.} \bibnamefont{Fischer}},
  \bibinfo{journal}{Phys. Rev. Lett.} \textbf{\bibinfo{volume}{103}},
  \bibinfo{pages}{052003} (\bibinfo{year}{2009}), \eprint{0904.2700}.

\bibitem[{\citenamefont{Fischer and Mueller}(2009)}]{Fischer:2009gk}
\bibinfo{author}{\bibfnamefont{C.~S.} \bibnamefont{Fischer}} \bibnamefont{and}
  \bibinfo{author}{\bibfnamefont{J.~A.} \bibnamefont{Mueller}},
  \bibinfo{journal}{Phys. Rev.} \textbf{\bibinfo{volume}{D80}},
  \bibinfo{pages}{074029} (\bibinfo{year}{2009}), \eprint{0908.0007}.

\bibitem[{\citenamefont{Braun et~al.}(2011)\citenamefont{Braun, Haas,
  Marhauser, and Pawlowski}}]{Braun:2009gm}
\bibinfo{author}{\bibfnamefont{J.}~\bibnamefont{Braun}},
  \bibinfo{author}{\bibfnamefont{L.~M.} \bibnamefont{Haas}},
  \bibinfo{author}{\bibfnamefont{F.}~\bibnamefont{Marhauser}},
  \bibnamefont{and} \bibinfo{author}{\bibfnamefont{J.~M.}
  \bibnamefont{Pawlowski}}, \bibinfo{journal}{Phys. Rev. Lett.}
  \textbf{\bibinfo{volume}{106}}, \bibinfo{pages}{022002}
  (\bibinfo{year}{2011}), \eprint{0908.0008}.

\bibitem[{\citenamefont{Fischer et~al.}(2010)\citenamefont{Fischer, Maas, and
  Muller}}]{Fischer:2010fx}
\bibinfo{author}{\bibfnamefont{C.~S.} \bibnamefont{Fischer}},
  \bibinfo{author}{\bibfnamefont{A.}~\bibnamefont{Maas}}, \bibnamefont{and}
  \bibinfo{author}{\bibfnamefont{J.~A.} \bibnamefont{Muller}},
  \bibinfo{journal}{Eur. Phys. J.} \textbf{\bibinfo{volume}{C68}},
  \bibinfo{pages}{165} (\bibinfo{year}{2010}), \eprint{1003.1960}.

\bibitem[{\citenamefont{Braun et~al.}(2010{\natexlab{c}})\citenamefont{Braun,
  Eichhorn, Gies, and Pawlowski}}]{Braun:2010cy}
\bibinfo{author}{\bibfnamefont{J.}~\bibnamefont{Braun}},
  \bibinfo{author}{\bibfnamefont{A.}~\bibnamefont{Eichhorn}},
  \bibinfo{author}{\bibfnamefont{H.}~\bibnamefont{Gies}}, \bibnamefont{and}
  \bibinfo{author}{\bibfnamefont{J.~M.} \bibnamefont{Pawlowski}},
  \bibinfo{journal}{Eur. Phys. J.} \textbf{\bibinfo{volume}{C70}},
  \bibinfo{pages}{689} (\bibinfo{year}{2010}{\natexlab{c}}),
  \eprint{1007.2619}.

\bibitem[{\citenamefont{Pawlowski}(2010)}]{Pawlowski:2010ht}
\bibinfo{author}{\bibfnamefont{J.~M.} \bibnamefont{Pawlowski}}
  (\bibinfo{year}{2010}), \eprint{1012.5075}.

\bibitem[{\citenamefont{Schaefer et~al.}(2010)\citenamefont{Schaefer, Wagner,
  and Wambach}}]{Schaefer:2009ui}
\bibinfo{author}{\bibfnamefont{B.-J.} \bibnamefont{Schaefer}},
  \bibinfo{author}{\bibfnamefont{M.}~\bibnamefont{Wagner}}, \bibnamefont{and}
  \bibinfo{author}{\bibfnamefont{J.}~\bibnamefont{Wambach}},
  \bibinfo{journal}{Phys. Rev.} \textbf{\bibinfo{volume}{D81}},
  \bibinfo{pages}{074013} (\bibinfo{year}{2010}), \eprint{0910.5628}.

\bibitem[{\citenamefont{Gattringer}(2006)}]{Gattringer:2006ci}
\bibinfo{author}{\bibfnamefont{C.}~\bibnamefont{Gattringer}},
  \bibinfo{journal}{Phys. Rev. Lett.} \textbf{\bibinfo{volume}{97}},
  \bibinfo{pages}{032003} (\bibinfo{year}{2006}), \eprint{hep-lat/0605018}.

\bibitem[{\citenamefont{Synatschke et~al.}(2007)\citenamefont{Synatschke, Wipf,
  and Wozar}}]{Synatschke:2007bz}
\bibinfo{author}{\bibfnamefont{F.}~\bibnamefont{Synatschke}},
  \bibinfo{author}{\bibfnamefont{A.}~\bibnamefont{Wipf}}, \bibnamefont{and}
  \bibinfo{author}{\bibfnamefont{C.}~\bibnamefont{Wozar}},
  \bibinfo{journal}{Phys. Rev.} \textbf{\bibinfo{volume}{D75}},
  \bibinfo{pages}{114003} (\bibinfo{year}{2007}), \eprint{hep-lat/0703018}.

\bibitem[{\citenamefont{Bilgici et~al.}(2008)\citenamefont{Bilgici, Bruckmann,
  Gattringer, and Hagen}}]{Bilgici:2008qy}
\bibinfo{author}{\bibfnamefont{E.}~\bibnamefont{Bilgici}},
  \bibinfo{author}{\bibfnamefont{F.}~\bibnamefont{Bruckmann}},
  \bibinfo{author}{\bibfnamefont{C.}~\bibnamefont{Gattringer}},
  \bibnamefont{and} \bibinfo{author}{\bibfnamefont{C.}~\bibnamefont{Hagen}},
  \bibinfo{journal}{Phys. Rev.} \textbf{\bibinfo{volume}{D77}},
  \bibinfo{pages}{094007} (\bibinfo{year}{2008}), \eprint{0801.4051}.

\bibitem[{\citenamefont{Kashiwa et~al.}(2009)\citenamefont{Kashiwa, Matsuzaki,
  Kouno, Sakai, and Yahiro}}]{Kashiwa:2008bq}
\bibinfo{author}{\bibfnamefont{K.}~\bibnamefont{Kashiwa}},
  \bibinfo{author}{\bibfnamefont{M.}~\bibnamefont{Matsuzaki}},
  \bibinfo{author}{\bibfnamefont{H.}~\bibnamefont{Kouno}},
  \bibinfo{author}{\bibfnamefont{Y.}~\bibnamefont{Sakai}}, \bibnamefont{and}
  \bibinfo{author}{\bibfnamefont{M.}~\bibnamefont{Yahiro}},
  \bibinfo{journal}{Phys. Rev.} \textbf{\bibinfo{volume}{D79}},
  \bibinfo{pages}{076008} (\bibinfo{year}{2009}), \eprint{0812.4747}.

\bibitem[{\citenamefont{Sakai et~al.}(2008)\citenamefont{Sakai, Kashiwa, Kouno,
  and Yahiro}}]{Sakai:2008py}
\bibinfo{author}{\bibfnamefont{Y.}~\bibnamefont{Sakai}},
  \bibinfo{author}{\bibfnamefont{K.}~\bibnamefont{Kashiwa}},
  \bibinfo{author}{\bibfnamefont{H.}~\bibnamefont{Kouno}}, \bibnamefont{and}
  \bibinfo{author}{\bibfnamefont{M.}~\bibnamefont{Yahiro}},
  \bibinfo{journal}{Phys. Rev.} \textbf{\bibinfo{volume}{D77}},
  \bibinfo{pages}{051901} (\bibinfo{year}{2008}), \eprint{0801.0034}.

\bibitem[{\citenamefont{Bilgici et~al.}(2010)}]{Bilgici:2009tx}
\bibinfo{author}{\bibfnamefont{E.}~\bibnamefont{Bilgici}} \bibnamefont{et~al.},
  \bibinfo{journal}{Few Body Syst.} \textbf{\bibinfo{volume}{47}},
  \bibinfo{pages}{125} (\bibinfo{year}{2010}), \eprint{0906.3957}.

\bibitem[{\citenamefont{Zhang et~al.}(2010)\citenamefont{Zhang, Bruckmann,
  Gattringer, Fodor, and Szabo}}]{Zhang:2010ui}
\bibinfo{author}{\bibfnamefont{B.}~\bibnamefont{Zhang}},
  \bibinfo{author}{\bibfnamefont{F.}~\bibnamefont{Bruckmann}},
  \bibinfo{author}{\bibfnamefont{C.}~\bibnamefont{Gattringer}},
  \bibinfo{author}{\bibfnamefont{Z.}~\bibnamefont{Fodor}}, \bibnamefont{and}
  \bibinfo{author}{\bibfnamefont{K.~K.} \bibnamefont{Szabo}}
  (\bibinfo{year}{2010}), \eprint{1012.2314}.

\bibitem[{\citenamefont{Mukherjee et~al.}(2010)\citenamefont{Mukherjee, Chen,
  and Huang}}]{Mukherjee:2010cp}
\bibinfo{author}{\bibfnamefont{T.~K.} \bibnamefont{Mukherjee}},
  \bibinfo{author}{\bibfnamefont{H.}~\bibnamefont{Chen}}, \bibnamefont{and}
  \bibinfo{author}{\bibfnamefont{M.}~\bibnamefont{Huang}},
  \bibinfo{journal}{Phys. Rev.} \textbf{\bibinfo{volume}{D82}},
  \bibinfo{pages}{034015} (\bibinfo{year}{2010}), \eprint{1005.2482}.

\bibitem[{\citenamefont{Gatto and Ruggieri}(2010)}]{Gatto:2010qs}
\bibinfo{author}{\bibfnamefont{R.}~\bibnamefont{Gatto}} \bibnamefont{and}
  \bibinfo{author}{\bibfnamefont{M.}~\bibnamefont{Ruggieri}},
  \bibinfo{journal}{Phys. Rev.} \textbf{\bibinfo{volume}{D82}},
  \bibinfo{pages}{054027} (\bibinfo{year}{2010}), \eprint{1007.0790}.

\bibitem[{\citenamefont{Kondo}(2010)}]{Kondo:2010ts}
\bibinfo{author}{\bibfnamefont{K.-I.} \bibnamefont{Kondo}},
  \bibinfo{journal}{Phys. Rev.} \textbf{\bibinfo{volume}{D82}},
  \bibinfo{pages}{065024} (\bibinfo{year}{2010}), \eprint{1005.0314}.

\bibitem[{\citenamefont{Wetterich}(1993)}]{Wetterich:1992yh}
\bibinfo{author}{\bibfnamefont{C.}~\bibnamefont{Wetterich}},
  \bibinfo{journal}{Phys. Lett.} \textbf{\bibinfo{volume}{B301}},
  \bibinfo{pages}{90} (\bibinfo{year}{1993}).

\bibitem[{\citenamefont{Litim and Pawlowski}(1998)}]{Litim:1998nf}
\bibinfo{author}{\bibfnamefont{D.~F.} \bibnamefont{Litim}} \bibnamefont{and}
  \bibinfo{author}{\bibfnamefont{J.~M.} \bibnamefont{Pawlowski}}
  (\bibinfo{year}{1998}), \eprint{hep-th/9901063}.

\bibitem[{\citenamefont{Bagnuls and Bervillier}(2001)}]{Bagnuls:2000ae}
\bibinfo{author}{\bibfnamefont{C.}~\bibnamefont{Bagnuls}} \bibnamefont{and}
  \bibinfo{author}{\bibfnamefont{C.}~\bibnamefont{Bervillier}},
  \bibinfo{journal}{Phys. Rept.} \textbf{\bibinfo{volume}{348}},
  \bibinfo{pages}{91} (\bibinfo{year}{2001}), \eprint{hep-th/0002034}.

\bibitem[{\citenamefont{Berges et~al.}(2002)\citenamefont{Berges, Tetradis, and
  Wetterich}}]{Berges:2000ew}
\bibinfo{author}{\bibfnamefont{J.}~\bibnamefont{Berges}},
  \bibinfo{author}{\bibfnamefont{N.}~\bibnamefont{Tetradis}}, \bibnamefont{and}
  \bibinfo{author}{\bibfnamefont{C.}~\bibnamefont{Wetterich}},
  \bibinfo{journal}{Phys. Rept.} \textbf{\bibinfo{volume}{363}},
  \bibinfo{pages}{223} (\bibinfo{year}{2002}), \eprint{hep-ph/0005122}.

\bibitem[{\citenamefont{Polonyi}(2003)}]{Polonyi:2001se}
\bibinfo{author}{\bibfnamefont{J.}~\bibnamefont{Polonyi}},
  \bibinfo{journal}{Central Eur. J. Phys.} \textbf{\bibinfo{volume}{1}},
  \bibinfo{pages}{1} (\bibinfo{year}{2003}), \eprint{hep-th/0110026}.

\bibitem[{\citenamefont{Delamotte et~al.}(2004)\citenamefont{Delamotte,
  Mouhanna, and Tissier}}]{Delamotte:2003dw}
\bibinfo{author}{\bibfnamefont{B.}~\bibnamefont{Delamotte}},
  \bibinfo{author}{\bibfnamefont{D.}~\bibnamefont{Mouhanna}}, \bibnamefont{and}
  \bibinfo{author}{\bibfnamefont{M.}~\bibnamefont{Tissier}},
  \bibinfo{journal}{Phys. Rev.} \textbf{\bibinfo{volume}{B69}},
  \bibinfo{pages}{134413} (\bibinfo{year}{2004}), \eprint{cond-mat/0309101}.

\bibitem[{\citenamefont{Pawlowski}(2007)}]{Pawlowski:2005xe}
\bibinfo{author}{\bibfnamefont{J.~M.} \bibnamefont{Pawlowski}},
  \bibinfo{journal}{Annals Phys.} \textbf{\bibinfo{volume}{322}},
  \bibinfo{pages}{2831} (\bibinfo{year}{2007}), \eprint{hep-th/0512261}.

\bibitem[{\citenamefont{Gies}(2006)}]{Gies:2006wv}
\bibinfo{author}{\bibfnamefont{H.}~\bibnamefont{Gies}} (\bibinfo{year}{2006}),
  \eprint{hep-ph/0611146}.

\bibitem[{\citenamefont{Delamotte}(2007)}]{Delamotte:2007pf}
\bibinfo{author}{\bibfnamefont{B.}~\bibnamefont{Delamotte}}
  (\bibinfo{year}{2007}), \eprint{cond-mat/0702365}.

\bibitem[{\citenamefont{Rosten}(2010)}]{Rosten:2010vm}
\bibinfo{author}{\bibfnamefont{O.~J.} \bibnamefont{Rosten}}
  (\bibinfo{year}{2010}), \eprint{1003.1366}.

\bibitem[{\citenamefont{Berges et~al.}(1999)\citenamefont{Berges, Jungnickel,
  and Wetterich}}]{Berges:1997eu}
\bibinfo{author}{\bibfnamefont{J.}~\bibnamefont{Berges}},
  \bibinfo{author}{\bibfnamefont{D.~U.} \bibnamefont{Jungnickel}},
  \bibnamefont{and}
  \bibinfo{author}{\bibfnamefont{C.}~\bibnamefont{Wetterich}},
  \bibinfo{journal}{Phys. Rev.} \textbf{\bibinfo{volume}{D59}},
  \bibinfo{pages}{034010} (\bibinfo{year}{1999}), \eprint{hep-ph/9705474}.

\bibitem[{\citenamefont{Braun}(2010)}]{Braun:2009si}
\bibinfo{author}{\bibfnamefont{J.}~\bibnamefont{Braun}},
  \bibinfo{journal}{Phys. Rev.} \textbf{\bibinfo{volume}{D81}},
  \bibinfo{pages}{016008} (\bibinfo{year}{2010}), \eprint{0908.1543}.

\bibitem[{\citenamefont{Gies et~al.}(2004)\citenamefont{Gies, Jaeckel, and
  Wetterich}}]{Gies:2003dp}
\bibinfo{author}{\bibfnamefont{H.}~\bibnamefont{Gies}},
  \bibinfo{author}{\bibfnamefont{J.}~\bibnamefont{Jaeckel}}, \bibnamefont{and}
  \bibinfo{author}{\bibfnamefont{C.}~\bibnamefont{Wetterich}},
  \bibinfo{journal}{Phys. Rev.} \textbf{\bibinfo{volume}{D69}},
  \bibinfo{pages}{105008} (\bibinfo{year}{2004}), \eprint{hep-ph/0312034}.

\bibitem[{\citenamefont{Fukano and Sannino}(2010)}]{Fukano:2010yv}
\bibinfo{author}{\bibfnamefont{H.~S.} \bibnamefont{Fukano}} \bibnamefont{and}
  \bibinfo{author}{\bibfnamefont{F.}~\bibnamefont{Sannino}},
  \bibinfo{journal}{Phys. Rev.} \textbf{\bibinfo{volume}{D82}},
  \bibinfo{pages}{035021} (\bibinfo{year}{2010}), \eprint{1005.3340}.

\bibitem[{\citenamefont{Hubbard}(1959)}]{Hubbard:1959ub}
\bibinfo{author}{\bibfnamefont{J.}~\bibnamefont{Hubbard}},
  \bibinfo{journal}{Phys. Rev. Lett.} \textbf{\bibinfo{volume}{3}},
  \bibinfo{pages}{77} (\bibinfo{year}{1959}).

\bibitem[{\citenamefont{Stratonovich}(1957)}]{Stratonovich}
\bibinfo{author}{\bibfnamefont{R.}~\bibnamefont{Stratonovich}},
  \bibinfo{journal}{Dokl. Akad. Nauk.} \textbf{\bibinfo{volume}{115}},
  \bibinfo{pages}{1097} (\bibinfo{year}{1957}).

\bibitem[{\citenamefont{Braun et~al.}(2010{\natexlab{d}})\citenamefont{Braun,
  Gies, and Scherer}}]{Braun:2010tt}
\bibinfo{author}{\bibfnamefont{J.}~\bibnamefont{Braun}},
  \bibinfo{author}{\bibfnamefont{H.}~\bibnamefont{Gies}}, \bibnamefont{and}
  \bibinfo{author}{\bibfnamefont{D.~D.} \bibnamefont{Scherer}}
  (\bibinfo{year}{2010}{\natexlab{d}}), \eprint{1011.1456}.

\bibitem[{\citenamefont{Dumitru et~al.}(2010)\citenamefont{Dumitru, Guo,
  Hidaka, Altes, and Pisarski}}]{Dumitru:2010mj}
\bibinfo{author}{\bibfnamefont{A.}~\bibnamefont{Dumitru}},
  \bibinfo{author}{\bibfnamefont{Y.}~\bibnamefont{Guo}},
  \bibinfo{author}{\bibfnamefont{Y.}~\bibnamefont{Hidaka}},
  \bibinfo{author}{\bibfnamefont{C.~P.~K.} \bibnamefont{Altes}},
  \bibnamefont{and} \bibinfo{author}{\bibfnamefont{R.~D.}
  \bibnamefont{Pisarski}} (\bibinfo{year}{2010}), \eprint{1011.3820}.

\bibitem[{\citenamefont{D'Elia et~al.}(2005)\citenamefont{D'Elia, Di~Giacomo,
  and Pica}}]{D'Elia:2005bv}
\bibinfo{author}{\bibfnamefont{M.}~\bibnamefont{D'Elia}},
  \bibinfo{author}{\bibfnamefont{A.}~\bibnamefont{Di~Giacomo}},
  \bibnamefont{and} \bibinfo{author}{\bibfnamefont{C.}~\bibnamefont{Pica}},
  \bibinfo{journal}{Phys. Rev.} \textbf{\bibinfo{volume}{D72}},
  \bibinfo{pages}{114510} (\bibinfo{year}{2005}), \eprint{hep-lat/0503030}.

\bibitem[{\citenamefont{Braun and Janot}()}]{BraunJanot}
\bibinfo{author}{\bibfnamefont{J.}~\bibnamefont{Braun}} \bibnamefont{and}
  \bibinfo{author}{\bibfnamefont{A.}~\bibnamefont{Janot}},
  \bibinfo{journal}{(work in progress)}.

\bibitem[{\citenamefont{Litim}(2000)}]{Litim:2000ci}
\bibinfo{author}{\bibfnamefont{D.~F.} \bibnamefont{Litim}},
  \bibinfo{journal}{Phys. Lett.} \textbf{\bibinfo{volume}{B486}},
  \bibinfo{pages}{92} (\bibinfo{year}{2000}), \eprint{hep-th/0005245}.

\bibitem[{\citenamefont{Litim}(2001{\natexlab{a}})}]{Litim:2001fd}
\bibinfo{author}{\bibfnamefont{D.~F.} \bibnamefont{Litim}},
  \bibinfo{journal}{Int. J. Mod. Phys.} \textbf{\bibinfo{volume}{A16}},
  \bibinfo{pages}{2081} (\bibinfo{year}{2001}{\natexlab{a}}),
  \eprint{hep-th/0104221}.

\bibitem[{\citenamefont{Litim}(2001{\natexlab{b}})}]{Litim:2001up}
\bibinfo{author}{\bibfnamefont{D.~F.} \bibnamefont{Litim}},
  \bibinfo{journal}{Phys. Rev.} \textbf{\bibinfo{volume}{D64}},
  \bibinfo{pages}{105007} (\bibinfo{year}{2001}{\natexlab{b}}),
  \eprint{hep-th/0103195}.

\bibitem[{\citenamefont{Litim and Pawlowski}(2006)}]{Litim:2006ag}
\bibinfo{author}{\bibfnamefont{D.~F.} \bibnamefont{Litim}} \bibnamefont{and}
  \bibinfo{author}{\bibfnamefont{J.~M.} \bibnamefont{Pawlowski}},
  \bibinfo{journal}{JHEP} \textbf{\bibinfo{volume}{11}}, \bibinfo{pages}{026}
  (\bibinfo{year}{2006}), \eprint{hep-th/0609122}.

\bibitem[{\citenamefont{Blaizot et~al.}(2007)\citenamefont{Blaizot, Ipp,
  Mendez-Galain, and Wschebor}}]{Blaizot:2006rj}
\bibinfo{author}{\bibfnamefont{J.-P.} \bibnamefont{Blaizot}},
  \bibinfo{author}{\bibfnamefont{A.}~\bibnamefont{Ipp}},
  \bibinfo{author}{\bibfnamefont{R.}~\bibnamefont{Mendez-Galain}},
  \bibnamefont{and} \bibinfo{author}{\bibfnamefont{N.}~\bibnamefont{Wschebor}},
  \bibinfo{journal}{Nucl. Phys.} \textbf{\bibinfo{volume}{A784}},
  \bibinfo{pages}{376} (\bibinfo{year}{2007}), \eprint{hep-ph/0610004}.

\end{thebibliography}
\end{document}